\documentclass[journal,twoside,web]{ieeecolor}
\usepackage{generic}
\usepackage{cite}
\usepackage{amsmath,amssymb,amsfonts}
\usepackage{algorithmic}
\usepackage{graphicx}
\usepackage{subfigure}
\usepackage{algorithm,algorithmic}
\usepackage{hyperref}
\usepackage{float}
\usepackage{textcomp}
\usepackage[flushleft]{threeparttable}
\usepackage{multirow}
\usepackage{array}
\newcolumntype{M}[1]{>{\centering\arraybackslash}m{#1}}

\def\BibTeX{{\rm B\kern-.05em{\sc i\kern-.025em b}\kern-.08em
    T\kern-.1667em\lower.7ex\hbox{E}\kern-.125emX}}
\begin{document}
\title{ViT-MDHGR: Cross-day Reliability and Agility in Dynamic Hand Gesture Prediction via HD-sEMG Signal Decoding 
}

\author{Qin Hu{$^{\dag}$}, Golara Ahmadi Azar{$^{\dag}$}, Alyson Fletcher, Sundeep Rangan, S. Farokh Atashzar{$^{*}$}, 
\thanks{This material is based upon work supported by the US National Science Foundation under Grant No \#2229697, \#2208189, \#2121391.}
\thanks{Hu, Rangan, and Atashzar are with the Department of Electrical and Computer Engineering, New York University (NYU), New York, NY, 11201 USA. Rangan is also the Director of NYU WIRELESS. Atashzar is also with the Department of Mechanical and Aerospace Engineering, Biomedical Engineering, NYU WIRELESS, and NYU Center for Urban Science and Progress (CUSP).}
\thanks{Azar and Fletcher are with the Department of Electrical and Computer Engineering, University of California Los Angeles (UCLA), Los Angeles, CA, 90095 USA. Fletcher is also with the Department of Statistics, Mathematics, and Computer Science, UCLA.}
\thanks{$^{\dag}$ Hu and Azar share the first authorship.\\
        $^{*}$ Corresponding author: {\tt\footnotesize f.atashzar@nyu.edu}.}
}

\maketitle

\begin{abstract}
Surface electromyography (sEMG) and high-density sEMG (HD-sEMG) biosignals have been extensively investigated for myoelectric control of prosthetic devices, neurorobotics, and more recently human-computer interfaces because of their capability for hand gesture recognition/prediction in a wearable and non-invasive manner. High intraday (same-day) performance has been reported. However, the interday performance (separating training and testing days) is substantially degraded due to the poor generalizability of conventional approaches over time, hindering the application of such techniques in real-life practices. There are limited recent studies on the feasibility of multi-day hand gesture recognition. The existing studies face a major challenge: the need for long sEMG epochs makes the corresponding neural interfaces impractical due to the induced delay in myoelectric control. This paper proposes a compact ViT-based network for multi-day dynamic hand gesture prediction. We tackle the main challenge as the proposed model only relies on very short HD-sEMG signal windows (i.e., 50 ms, accounting for only one-sixth of the convention for real-time myoelectric implementation), boosting agility and responsiveness. Our proposed model can predict 11 dynamic gestures for 20 subjects with an average accuracy of over 71\% on the testing day, 3-25 days after training. Moreover, when calibrated on just a small portion of data from the testing day, the proposed model can achieve over 92\% accuracy by retraining less than 10\% of the parameters for computational efficiency.
\end{abstract}


\section{Introduction}
\label{sec:intro}
The initiation of voluntary human motion starts from the central nervous system, which sends synaptic inputs to the motor neurons in the spinal cord \cite{Feher2012-ia}. These inputs are then transduced into forces and transmitted through action potentials by motor units that comprise muscle fibers. Surface electromyography (sEMG) is the cumulative sum of motor unit action potentials \cite{Del_Vecchio2020-if}. It is measured and recorded from the skin surface during muscle activities. The early research of sEMG has the main interest in clinical diagnosis and biomedical applications, such as rehabilitation and assistive technologies (e.g., prosthetic control) and ergonomics \cite{Raez2006-ro, Bi2019-up}. Nowadays, sEMG, which is the control signal in human-computer interaction, has a wide range of applications, such as in augmented reality and virtual reality (for hands-free control to counterbalance network delay) \cite{Zhang2022-ch} and sports science (for performance measurement and optimization) \cite{Ergeneci2022-eq}. The increasing research interest in sEMG is due to its non-invasiveness, wearability, and potential for real-time control. Over the past two decades, high-density sEMG (HD-sEMG) has enhanced the performance of these applications by capturing signals with high spatial resolution information on muscle activities.

sEMG-based hand gesture recognition (HGR) is achieved through pattern recognition (PR), which mainly includes two progressive steps: feature extraction and classification. Machine learning (ML) and deep learning (DL) are the two basic processing methods in sEMG PR \cite{Kumar2022-ux}. Classical ML relies on researchers' expertise and feature engineering for feature extraction and models such as Support Vector Machine (SVM) for classification \cite{Shehata2021-ns}. DL methods such as Convolutional Neural Networks (CNNs) automatically extract various temporal, spectral, and spatial features from sEMG through their hierarchical architectures \cite{Holobar2021-hb}. Conventionally, these methods trained and tested on sEMG signals collected on the same day only provide short-term performance but underestimate long-term practicality due to the disregard for the effect of sensor misplacement, sensor displacement, and day-to-day variation in human neurophysiology and skin conductivity \cite{Chowdhury2013-wf}. The performance of a previously trained model can drop by 55\% if used on another day \cite{Botros2022-rd}. To enhance the scalability and feasibility of such wearable neural interfaces for human-robot interaction and control, temporal generalizability should be addressed by securing high interday reliability.

Researchers have started investigating multi-day HGR based on sEMG to improve the longitudinal performance of these control systems within the last five years. These works can also be categorized into traditional ML and DL methods. Traditional ML relies on training on the extracted features from multiple days to capture sEMG variation across days and secure high interday performance without calibration. Adaptive ML approaches such as Adaptive LDA (ALDA) can adapt to new data from the testing day by updating its parameters to improve interday performance. However, at root, the main hurdles of these traditional ML methods are (1) the simplicity of the extracted features that are incompetent to capture cross-day characteristics in sEMG and generalize the use of multi-day HGR systems over a large number of subjects and (2) the computational inefficiency of the feature extraction processes, such that extracting 4th-order autoregression from highly spatiotemporal HD-sEMG collected from 256 electrodes can be more than half a second; these processes will be even less efficient on wearable devices with limited power. Researchers favor DL approaches in sEMG-based research due to their strong feature extraction power, which relaxes the need for expertise in sEMG compared to feature-engineering-based conventional ML. Adaptive DL methods have used domain adaptation (DA) to calibrate their previously trained models from one day to another to enhance the interday performance. Although most of the existing DL efforts have reduced the pre-processing time by directly feeding their deepnets with raw sEMG signals, their methods did not minimize the induced myoelectric control delay by using window sizes from 150 ms to 300 ms and sEMG signals from the plateau phase.

Regardless of the approaches, the limitations of the current literature can be summarized into (1) time-consuming feature extraction, (2) large window sizes ranging from 150 to 750 ms, and (3) the use of plateau-phase sEMG, neglecting the rich information during the dynamic transient phase when gesture-associated motor unit recruitment occurs. Other constraints of existing literature include but are not limited to (1) the low number of subjects (i.e., $<$20), which makes a control system not representative enough to general users, (2) the exclusion of dynamic gestures, which makes the system unrealistic since hand gestures are naturally dynamic rather than static in real-world applications, and (3) training on data only from multiple consecutive days (where the potential changes in sEMG can be minimized), which is impractical due to the progressive performance degradation when the interval between training and testing days is large.

In this paper, we propose a multi-day dynamic hand gesture prediction method through decoding HD-sEMG signals to overcome the above limitations. We aim for a seamless (i.e., real-time-ready) and realistic hand gesture prediction system based on a deep vision transform structure. The restriction of control delay is lifted by investigating 11 dynamic gestures and using a small window size of 50 ms. Compared to static gestures, dynamic gestures are more natural and realistic, thus more complex to detect due to the dynamicity of neurophysiology during such tasks. It should be noted that the 50 ms window size is only one-sixth of the maximum window size qualified for real-time implementation \cite{Englehart2003-iq} and the smallest size, to the best of our knowledge, in multi-day HGR research. Leveraging the power of the Multi-headed Self-attention (MSA) mechanism in a vision transformer (ViT), the proposed model can predict dynamic 11 gestures with an average accuracy of over 71\% across 20 subjects on the testing day, 3-25 days after training. When calibrated on minimum data from the testing day (i.e., one or two repetitions), the proposed model can achieve over 92\% accuracy by retraining only 8.8\% of the parameters, almost regaining the intraday accuracy, for which the proposed model is trained and tested on the same day.

The \textit{main contributions} of this paper to the research of multi-day HGR are as follows:
\begin{itemize}
    \item This paper proposes the first ViT-based network to improve the multi-day performance of dynamic hand gesture prediction by capturing the temporal relationships within each HD-sEMG window through attention mechanisms.
    \item 50 ms is the shortest window size used in multi-day HGR, minimizing the latency of myoelectric control to the best of our knowledge. This window size is only one-sixth of the requirement for real-time implementation, preparing the proposed model for seamless real-life practices.
    \item This paper investigates the minimum calibration data needed from the testing day to maintain the performance of the proposed model trained on a previous day and reused on the testing day. This investigation, which falls under the umbrella of few-shot learning, is for the first time in the research of sEMG-based multi-day HGR when only two days of data are available.
\end{itemize}

The rest of the paper is organized as follows: the Prior Works section summarizes the previous studies in single-day and multi-day HGR. The Dataset and Pre-processing section describes the acquisition and pre-processing of the datasets. The model architecture, the model evaluation protocols, and the statistical test are introduced in the Methods section. The Results section presents the model performance, followed by the comparative study section that compares our proposed method with the state-of-the-art approaches when solving the same multi-day HGR task. Finally, the Conclusion section summarizes this paper.

\section{Prior Works}
\label{sec:prior}
\setlength\tabcolsep{2pt}
\begin{table*}[tbh]
    \small
    \centering
    \caption{Comparison between the proposed model with the state-of-the-art efforts in sEMG-based multi-day HGR.}
    \label{tab:liter-review}
    \begin{tabular}{|M{0.8cm}|M{1cm}|M{1.2cm}|M{1cm}|M{1cm}|M{0.9cm}|M{1.1cm}|M{1.3cm}|M{4.6cm}|M{3.6cm}|}
        \hline
        \textbf{Paper} & \textbf{\# Subs} & \textbf{\# Moves} & \textbf{\# Reps} & \textbf{\# Days} & \textbf{Rep Len} & \textbf{Win Len} & \textbf{Signal Type} & \textbf{Method} & \textbf{Interday Acc} \\
        \hline\hline
        \cite{Zia_ur_Rehman2018-mk} & 10 & 10+rest & 4 & 7c & 3 s & 200 ms & plateau & feature engineering+AE & 66-80\%\\
        \hline
        \cite{Wang2019-qd} & 6 & 12+rest & 1 & 10c & 5 s & 256 ms & plateau & feature engineering+LDA & 72.3\%\\
        \hline
        \cite{Meng2022-hn} & 10 & 11+rest & 5 & 10 & 4 s & 200 ms & complete & feature engineering+(KNN,LDA) & 86.61\%\\
        \hline
        \cite{Jiang2022-di} & 20 & 10+rest & 6 & 2 & 1 s & 500 ms & dynamic & feature engineering+SVM & 92.2\%\\
        \hline
        \cite{Waris2019-jx} & 10 & 10+rest & 4 & 7c & 3 s & 160 ms & plateau & feature engineering+ANN & 76.2-85.6\%\\
        \hline
        \cite{Fang2021-ro} & 6 & 12+rest & 1 & 10c & 5 s & 256 ms & plateau & (feature engineering,raw)+CNN & 83.42\%\\
        \hline
        \cite{Zanghieri2020-tf} & 10 & 7 & 12 & 5c & 6 s & 150 ms & complete & raw+TCN & 49.4\%\\
        \hline
        \cite{Qureshi2022-wl} & 8 & 10+rest & 4 & 7c & 5 s & 200 ms & complete & mel spectrogram+CNN & 65.88-88.73\%\\
        \hline
        \cite{Ketyko2019-rm} & 10 & 8 & 10 & 2 & 1 s & 300 ms & plateau & raw+RNN & 54.6\% (w/o cal); \hspace{1cm}83.8 (w/ cal)\\
        \hline
        \cite{Du2017-il} & 10 & 8 & 10 & 2 & 1 s & 150 ms & plateau & raw+CNN+AdaBN & 63.3\%\\
        \hline
        \cite{Cote-Allard2020-nr} & 20 & 10+rest & 3 & 3 & 5 s & 150 ms & dynamic & feature engineering+SCADANN & 53.08\% (w/o cal); \hspace{1cm}55.69\% (w/ cal)\\
        \hline
        \textbf{ours} & \textbf{20} & \textbf{11} & \textbf{6} & \textbf{2} & \textbf{1 s} & \textbf{50 ms} & \textbf{dynamic} & \textbf{ViT} & \textbf{71.34\% (w/o cal);} \hspace{1cm}\textbf{88.87\% (w/ 1 rep cal);} \hspace{1cm}\textbf{92.25\% (w/ 2 reps cal)}\\
        \hline
    \end{tabular}
    \begin{tablenotes}
        \scriptsize
        \item Note: \#: Number; Subs: Subjects; Rep: Repetition; Len: Length; Win: Window; Acc: Accuracy; s: Second; ms: Millisecond; c: Consecutive; w/o: Without; w/: With; cal: Calibration; KNN: K-Nearest Neighbors; ANN: Artificial Neural Network; AdaBN: Adaptive Batch Normalization; SCADANN: Self-Calibrating Asynchronous Domain Adversarial Neural Network.
    \end{tablenotes}
\end{table*}

\subsubsection*{Single-day HGR}
For single-day hand gesture recognition/prediction, a model is trained and tested on sEMG signals collected from the same day. Classical ML relies on researchers' expertise and feature engineering for feature extraction and models, such as LDA \cite{Fang2017-fd,Raurale2020-ma}, SVM \cite{Tavakoli2018-hb}, and K-nearest Neighbor (KNN) \cite{Amsuess2016-hp}, for classification. CNNs \cite{Lee2020-tb}, Recurrent Neural Networks (RNNs) \cite{Sun2022-ib}, and other DL methods \cite{Liu2023-mc} automatically extract various temporal, spectral, and spatial features from sEMG through their hierarchical architectures.

Although researchers have reported high performance on a large number of classified gestures, the short-term (intraday) performance cannot be translated into long-term (interday) performance due to the changes in the characteristics of sEMG signals over time caused by electrode artifacts, electrode displacement, and electrode misplacement. Such artifacts mainly include (a) stochastic electromagnetic noises (such as fluorescent noise,  power-line noise, and those by nearby electronics), (b) signal deterioration due to degraded electrode skin contact impedance (due to hair blockage and sweat), and (c) changes in capacitive coupling \cite{Stachaczyk2020-je}. Electrode displacement results from electrode shift on the skin surface \cite{Boschmann2014-nc}, while electrode misplacement happens due to imprecise electrode positioning \cite{Moin2018-xb}. Other factors include but are not limited to natural sEMG variation over time and the muscle contraction effort of subjects \cite{Phinyomark2020-jb}. Due to the lack of robustness of the existing models to the sources of signal variation, commercial sEMG-PR-based control systems (e.g., myoelectric prostheses) are currently limited on the market \cite{Jiang2023-sd, Prakash2021-gq}. Therefore, it is crucial to address the day-to-day reliability by proposing algorithms that can generalized to sEMG collected from multiple days, enhancing the scalability and feasibility of such wearable neural interfaces for human-robot interaction and control.

\subsubsection*{Multi-day HGR}
Researchers have investigated multi-day HGR, where a model is trained on the previous day(s) and reused on the testing day with or without being calibrated on the data from the testing day. These works can also be categorized into conventional ML and DL methods.
Existing research based on conventional ML \cite{Zia_ur_Rehman2018-mk,Wang2019-qd,Meng2022-hn,Jiang2022-di,Waris2019-jx} relies on manually extracting commonly used temporal (e.g., Hudgin's time-domain features \cite{Hudgins1993-rq} and autoregression coefficients), spectral (e.g., median frequency and spectral entropy) features. The extracted features are often optimized using dimensionality reduction techniques (e.g., principal component analysis) to achieve computational efficiency and prevent overfitting. A traditional ML classifier (e.g., SVM) is then trained on the features extracted from sEMG signals collected from the previous day(s) and tested on the new day. One of the problems of the traditional ML methods based on feature engineering is the lack of adaptability to the data from the new day. Adaptive LDA (ALDA) is one of the most commonly used adaptive ML approaches in this category \cite{Gu2018-oo,Jiang2018-rl,Botros2022-stability,Hohne2015-covariate}. ALDA can be calibrated on the new day's data by updating its mean and covariance matrices based on the same parameters from previous and new days to capture the interday sEMG variation to improve interday performance. 

Existing DL works in multi-day HGR have developed deepnets (e.g., CNNs and autoencoders or AEs) to automatically extract HGR-related features from raw sEMG signals in the time domain \cite{Fang2021-ro,Zanghieri2020-tf,Zia_Ur_Rehman2018-bp}, or from spectrograms in the time-and-frequency domain \cite{Qureshi2022-wl}. Domain adaptation (DA) is the adaptive technique for DL methods to calibrate their previously trained models from one day to another to enhance the interday performance. DA aims to develop a discriminative predictor on the data from the source domain and then to adapt the predictor to the data from the target domain, which is different but related to the source domain, possibly achieving high performance on the testing day. One way to apply DA is through transfer learning using labeled data from the testing day \cite{Ketyko2019-rm}. The proposed deepnet consists of a DA layer followed by a classifier. In the pre-training stage, only the classifier is trained on sEMG collected before the testing day (source domain) while the DA layer is frozen. During the DA stage, the classifier is frozen at its pre-trained stage while the DA layer is trained on sEMG collected on the testing day (target domain). The other way is to reduce the domain divergence between the labeled sEMG (in the source domain) and the unlabeled sEMG (in the target domain) by progressively updating their proposed deepnets using the unlabeled sEMG based on techniques such as pseudo-labels generating heuristic and Adaptive Batch Normalization \cite{Du2017-il,Cote-Allard2020-nr,Wu2023-cs}. The comparison between our paper and the existing state-of-the-art efforts in sEMG-based multi-day HGR is summarized in Table \ref{tab:liter-review}.

\section{Dataset and Pre-processing}
\label{sec:db}
\subsection{Dataset}
\label{subsec:data}
\begin{figure}
\center
\subfigure[]{\includegraphics[width = 0.2\linewidth]{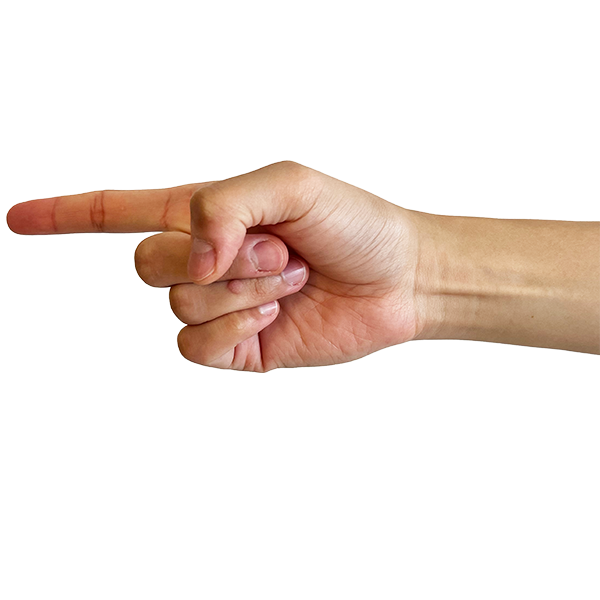}}
\subfigure[]{\includegraphics[width = 0.2\linewidth]{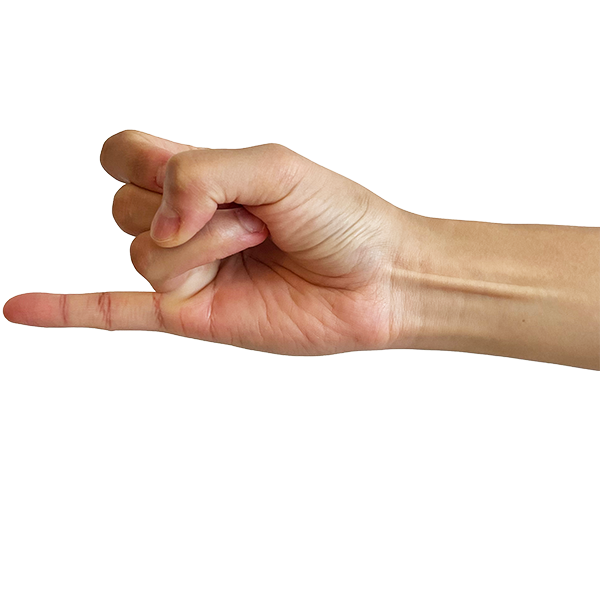}} 
\subfigure[]{\includegraphics[width = 0.2\linewidth]{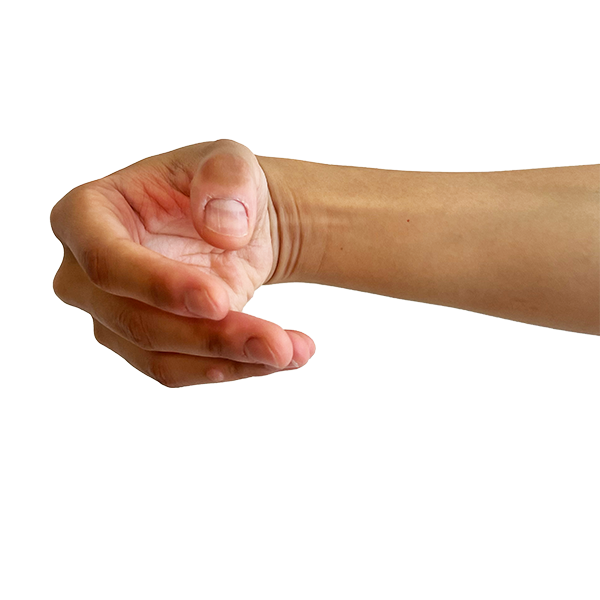}}
\subfigure[]{\includegraphics[width = 0.2\linewidth]{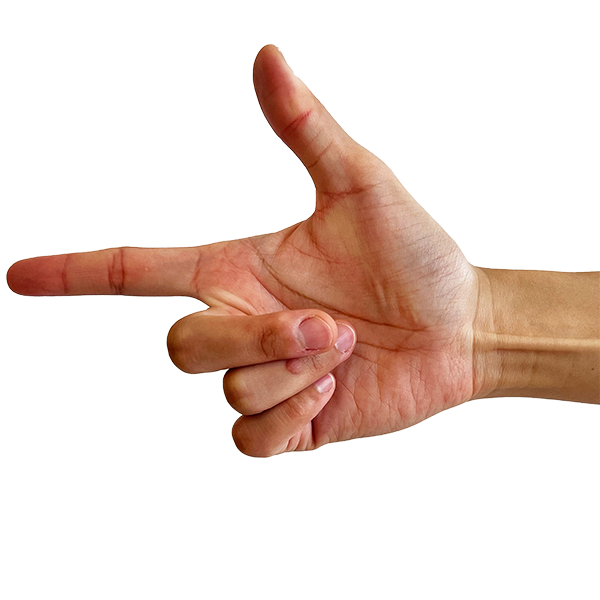}}\\
\subfigure[]{\includegraphics[width = 0.2\linewidth]{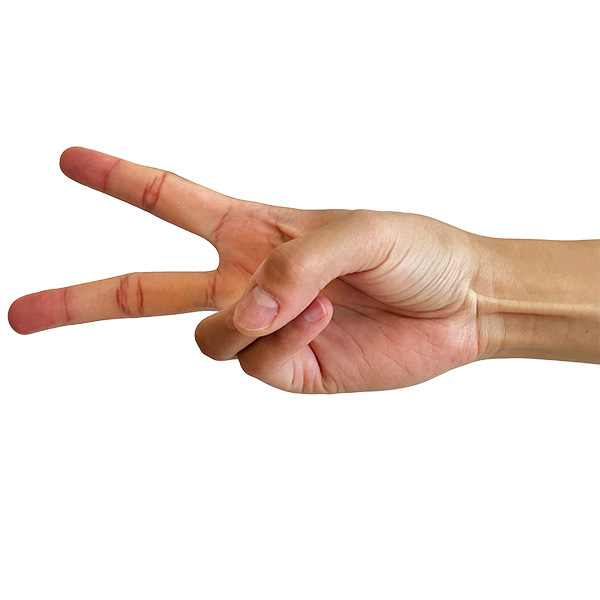}}
\subfigure[]{\includegraphics[width = 0.2\linewidth]{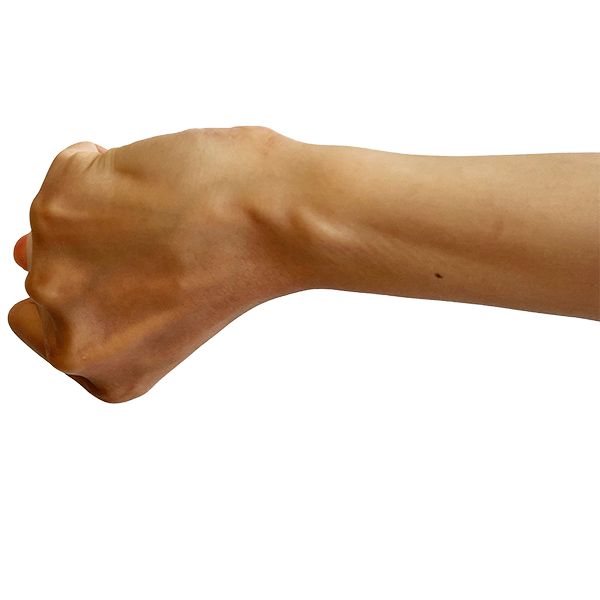}}
\subfigure[]{\includegraphics[width = 0.2\linewidth]{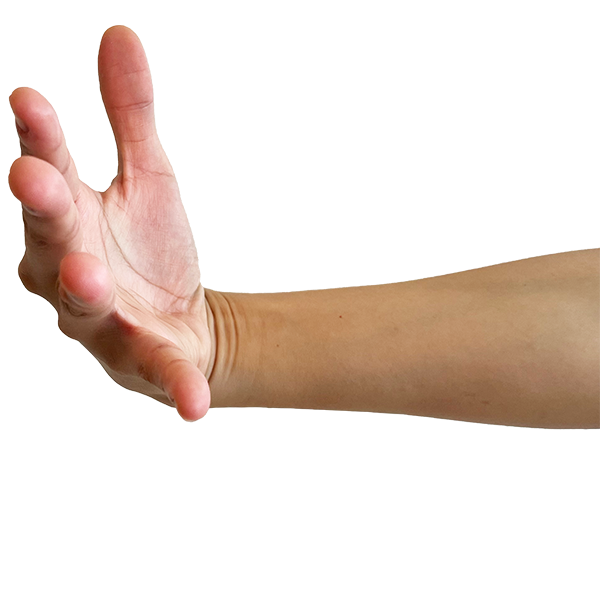}} 
\subfigure[]{\includegraphics[width = 0.2\linewidth]{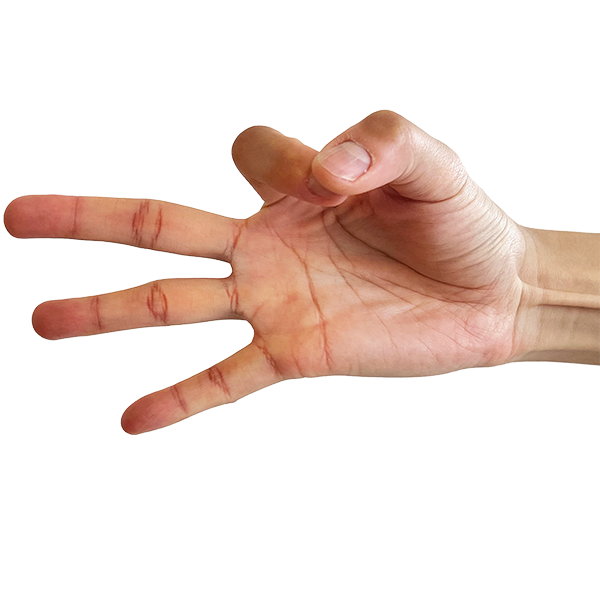}}\\ 
\subfigure[]{\includegraphics[width = 0.2\linewidth]{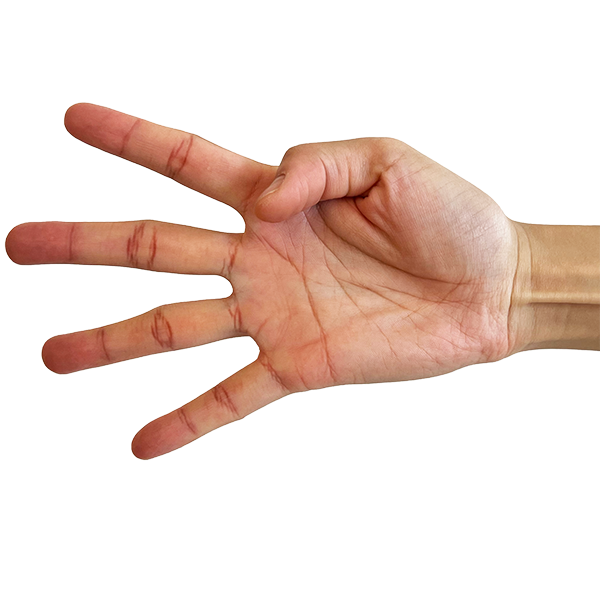}} 
\subfigure[]{\includegraphics[width = 0.2\linewidth]{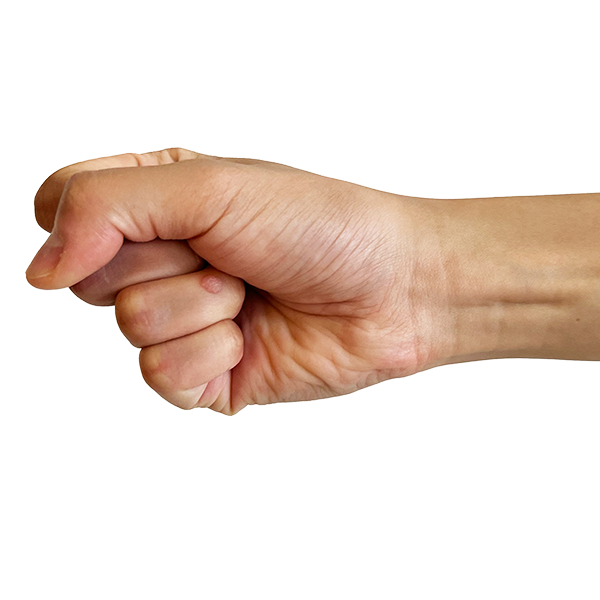}} 
\subfigure[]{\includegraphics[width = 0.2\linewidth]{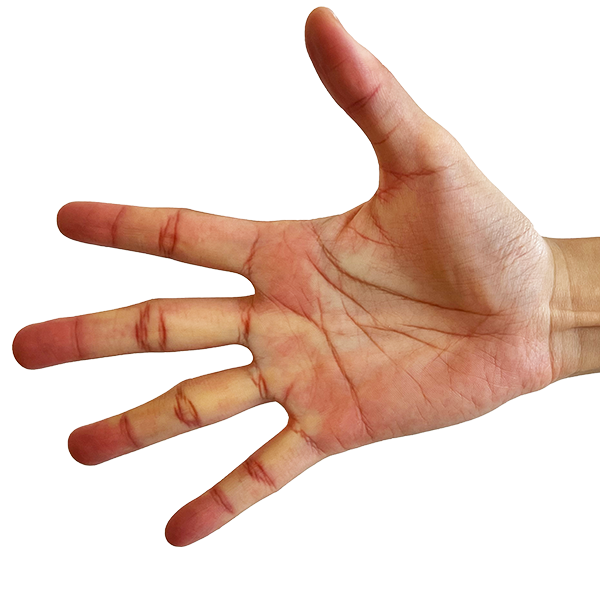}}
\caption{11 hand gestures that are common to all 20 subjects in the Hyser PR dataset are demonstrated. These gestures are the following: (a) index finger extension (IFE), (b) little finger extension (LFE), (c) wrist flexion (WF), (d) extension of thumb and index fingers (ETIF), (e) extension of index and middle fingers (EIMF), (f) wrist supination combined with hand close (WSHC), (g) wrist flexion combined with hand open (WFHO), (h) extension of middle, ring and little fingers (EMRLF), (i) extension of index, middle, ring and little fingers (EIMRLF), (j) hand close (HC), and (k) hand open (HO).}
\label{fig:gestures}
\end{figure}

We employ the PR dataset in a publicly available HD-sEMG database, referred to as  “Hyser” \cite{Jiang2021-hy}. This dataset includes 34 hand and finger gestures recorded from 20 intact subjects (12 male and 8 female with an average age of 26.5). In order to incorporate two-day data from all 20 subjects for statistical analysis, 11 gestures that are common to all subjects have been included in this study. Fig. \ref{fig:gestures} shows these gestures.

HD-sEMG signals were recorded using four $8 \times 8$ electrode array patches (256 sensors in total) by attaching two patches of sensors at each of the forearm sides (extensor and flexor) with a sampling rate of 2048 Hz. Each subject was instructed to perform two trials for each gesture, each trial containing three dynamic tasks of one-second duration from the resting state to the desired gesture and one maintenance task of holding that gesture for four seconds. In this study, we focus only on the dynamic tasks, resulting in six repetitions of one-second duration per gesture for each individual subject. Data were recorded from two different days with a between-day interval of 3 to 25 days, hereafter referred to as Day 1 (first day of recording) and Day 2 (second day of recording).

\subsection{Data Pre-processing}
``Hyser" pre-processed their PR dataset by applying a 10-500 Hz 8th-order Butterworth band-pass filter, followed by a notch filter that attenuates power-line interference at 1st-8th harmonics of 50 Hz. These pre-processed signals are further filtered by an 8th-order Butterworth low-pass filter at 200 Hz. Then, the first 250 ms of the signals corresponding to the reaction time are removed before windowing. Next, each signal is segmented into 50 ms windows with a stride of 10 ms, complying with the real-time requirements. More specifically, this window length corresponds to one-sixth of the maximum window length allowed for real-time analysis. To the best of our knowledge, this is the shortest window length used for multi-day gesture detection tasks using sEMG signals. In order to minimize all possible delays, no further pre-processing is performed on the windowed data. The input data shape is adjusted according to the model at hand. For the proposed ViT model, we use the input shape of $100 \times 4 \times 8 \times 8$ corresponding to $\text{window length} \times \text{\# electrode grids} \times \text{electrode width} \times \text{electrode length}$. For RNN models (described in section \ref{sec:comparative}), we use the input shape of $100 \times 256$ corresponding to $\text{window length} \times \text{\# electrodes}$.

\section{Methods}
\label{sec:methods}
\subsection{Model Architecture}
\label{subsec:model}
\begin{figure*}[tbh]
    \centering
    \includegraphics[width=\textwidth]{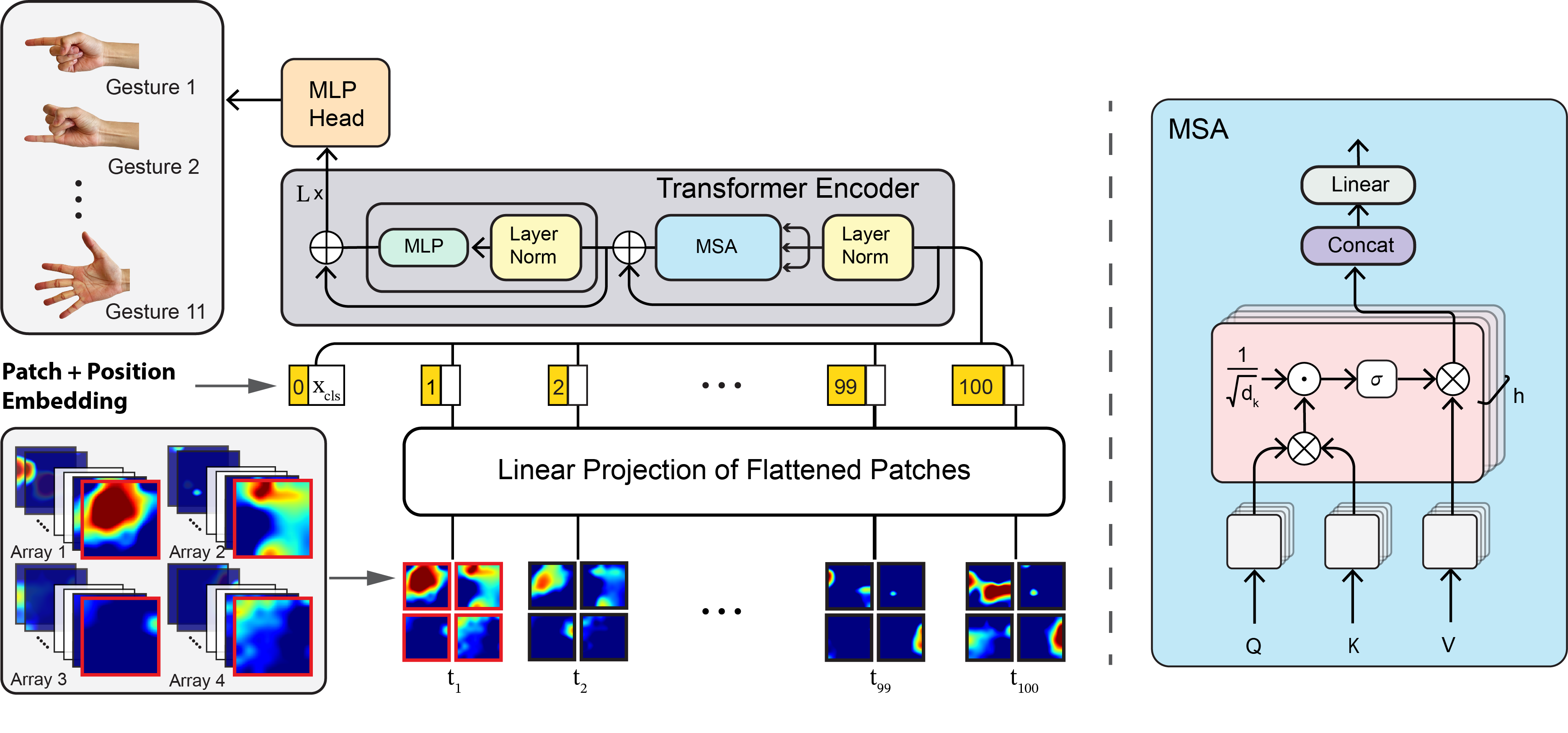}
    \caption{Model architecture. The entire model is trained during the pre-training stage, while only the Linear Projection layer (the white rectangle connected to HD-sEMG heatmaps), which accounts for only 8.8\% model complexity, is retrained. $\sigma$ denotes the $\mathrm{softmax}$ function in MSA.}
    \label{fig:arch}
\end{figure*}

ViTs have attracted extensive attention in image classification, challenging the dominance of CNNs. They introduce fewer inductive biases into the architectures than CNNs, which allows them to capture global contextual information by dividing each image into non-overlapping patches. Attention mechanisms enable ViTs to focus on relevant patches and establish long-range dependencies to learn complex patch relationships, capturing important visual patterns. Because of the patch extraction, ViTs are also scalable for high-resolution tasks that require fine-grained details. Position embedding preserves the spatial arrangement of the patches, whereas the spatial information may be lost due to the pooling layers of CNNs \cite{dosovitskiy20, Arnab2021-et}. sEMG-based HGR can benefit from the above advantages of ViTs, but ViT-based HGR methods using sEMG or HD-sEMG have been rarely studied in the literature, neglecting the temporal relationship between signals at any two timestamps. Montazerin et al. \cite{Montazerin2023-yr} just published the first ViT-based model for HD-sEMG-based HGR with intraday evaluation. In this paper, for the first time, we propose a compact deep neural network with a ViT backbone in multi-day dynamic hand gesture prediction from HD-sEMG signals, named ViT-MDHGR. The proposed model captures cross-day features by learning the relationships between HD-sEMG signals at any two timestamps within a window. The four components of the proposed model (patch embedding, position embedding, transformer encoder, and Multi-Layer Perceptron or MLP head) are introduced below. We follow most of the notations from the original ViT paper \cite{dosovitskiy20}.

\subsubsection*{Patch Embedding}
The raw input of each HD-sEMG window can be represented as a tensor $x \in \mathbb{R}^{T \times N_{g} \times H \times W}$, where $T$ is the window length, $N_{g}$ is the number of electrode grids, and $H \times W$ is the shape of each electrode grid. In our experiment, $T=100$, $N_g=4$, and $H\times W = 8 \times 8$.  As we aim to learn the relationships between the signals at any two timestamps of an HD-sEMG window, we consider the signals from the electrode grids at timestamp, $x[i,:,:,:]$, $i=1,\ldots,T$, as a patch. Therefore, we reshape an HD-sEMG window input $x$ into a sequence of flattened patches $x_p \in \mathbb{R}^{T \times N_{g}HW}$. A trainable linear projection layer maps these flattened patches at each time stamp to size $D$ latent vectors. The output of the projection is the patch embeddings. Then we prepend a class token, a vector of learnable embeddings $x_{cls} \in \mathbb{R}^D$, to the patch embeddings, turning the shape of patch embeddings to $\mathbb{R}^{(T+1) \times D}$. The state of the class token at the output of the transformer encoder is referred to as the classification head, which serves as the representation of an HD-sEMG window to be classified as different gestures.

\subsubsection*{Position Embedding}
Learnable position embeddings, $p \in \mathbb{R}^{(T+1) \times D}$, are added to the patch embeddings (including the class token) to retain the position information of the patches. Since we are interested in the patch relationships along the time axis and consider the input as a sequence rather than a grid of patches, we choose one-dimensional position embeddings. The position embeddings are initialized from a standard normal distribution. The input of the transformer encoder is:
\begin{equation}
    z_0 = [x_{cls}, x_p^1E, x_p^2E, \ldots, x_p^TE] + p\text{,}
\end{equation}
where $E \in \mathbb{R}^{N_{g}HW \times D}$ is the matrix for the linear projection.  

\subsubsection*{Transformer Encoder}
The transformer encoder consists of multiple layers of MSA and MLP blocks. Pre-norm happens before each block, where Layer Normalization (LN) is applied to the block input to estimate the normalization statistics and residual connections after (by adding the block input to the output) to improve model convergence. The outputs of an MSA block and an MLP block on layer $\ell$ are (\ref{eq:msa}) and (\ref{eq:mlp}), respectively.
\begin{align}
    z'_\ell &= MSA(LN(z'_{\ell-1})) + z'_{\ell-1} \label{eq:msa} \\
    z_{\ell} &= MLP(LN(z'_{\ell})) + z'_{\ell}\label{eq:mlp}
\end{align}

MSA has a building block of the standard \textbf{qkv} self-attention \cite{Vaswani2017-bv}, where the sequence of patches $z_0$ is linearly projected separately into $Q$ (queries), $K$ (keys), and $V$ (values), all with the same dimension of $h \times (T+1) \times d_k$, $h$ denotes the number of heads and $d_k$ is the head dimension. The attention weights $A$ are the similarities between two patches of the sequence and are calculated as
\begin{equation}
    A = \mathrm{softmax}(\frac{QK^\intercal}{\sqrt{d_k}}) \text{.}
\end{equation}

The scaling factor $\sqrt{d_k}$ ensures the dot product of $Q$ and $K$ will not be a large number. The $\mathrm{softmax}$ function converts the scaled dot product to the range between 0 and 1, indicating the total attention paid to $V$ sums up to 1. Therefore, the self-attention operation for each head is 

\begin{equation}
    MSA(Q,K,V) = \mathrm{softmax}(\frac{QK^\intercal}{\sqrt{d_k}})V \text{.}
\end{equation}

The self-attention operations of $h$ heads are run in parallel. The outputs of all the heads are concatenated and projected to form the input of the MLP block after the residual connection.

An MLP block consists of two linear layers with Gaussian Error Linear Unit (GELU) activation and a dropout in between. The output of an MLP block matches the dimension of the latent space $D$.

\subsubsection*{MLP Head}
The MLP head is a linear classifier that predicts the classification head $z_L^0$, the class token from the last layer $L$ of the transformer encoder, into different gestures $\hat{y}$ as follows:

\begin{equation}
    \hat{y} = \mathrm{softmax}(FC(LN(z_L^0))) \text{,}
\end{equation}

\noindent where $FC$ denotes ``fully connected", mapping $z_L^0$ from dimension $D$ to the total number of predicted gestures (i.e., 11 in this paper).

The proposed model is trained for a maximum of 200 epochs with a batch size of 32 for all the experiments, including pre-training and calibration. We use Adam as the optimizer with an adaptive learning rate of 0.001, which will be reduced to half at Epoch 40 and 80. Early stopping is employed with patience 40, such that the proposed model will stop training if the validation accuracy does not improve for 40 consecutive epochs.

\subsection{Model Evaluation Protocols}
\label{subsec:protocols}
\begin{figure}[thb]
    \centering
    \includegraphics[width=\linewidth]{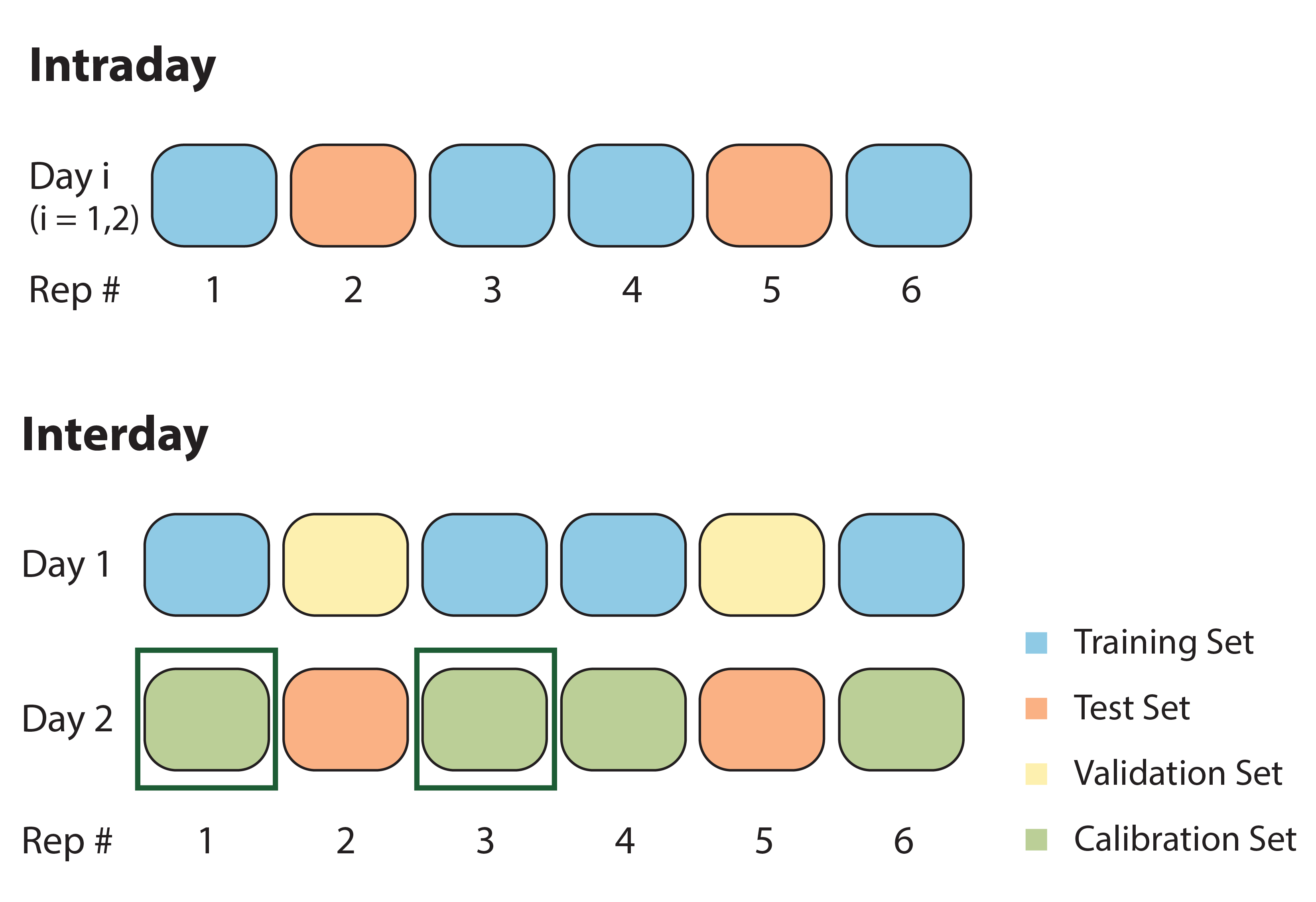}
    \caption{Visualization of training/training-calibration strategy of model evaluation for intraday (upper figure) and interday (lower figure) performances. Each colored rectangle with rounded corners represents a repetition. The two dark green rectangles show an example of 2 Reps Calibration on the repetitions (1 and 3) from Day 2.}
    \label{fig:strategies}
\end{figure}

This paper evaluates the proposed model for intraday and interday performance. The intraday performance serves as a benchmark to be compared with the interday performance to see (1) the acceptable performance degradation of our model trained on one day and used on another without calibration and (2) the slim performance gap between the intraday performance and the interday performance when calibrating our model on limited data from the testing day.

\subsubsection*{Intraday Performance} To evaluate the intraday performance of our proposed model on Day 1 or Day 2, we use the four repetitions (1, 3, 4, and 6) and the remaining two repetitions (2 and 5) of the same day for training and testing, respectively.

\subsubsection*{Interday Performance} We use the signals from Day 1 for training and validation and Day 2 for calibration and testing. As this paper leverages transfer learning to learn the HGR-related pre-knowledge on Day 1 and transfer this knowledge to each subject to predict gestures on Day 2, the evaluation of interday performance consists of two stages: pre-training and calibration stages.

This paper investigates two strategies in the pre-training stage: (1) pre-training our proposed model on the data from individual subjects, named ``Pre-trained on Individuals", and (2) pre-training the proposed model on the data pooled from all 20 subjects, named ``Pre-trained on All". The first pre-trained strategy will result in 20 pre-trained models, one for each subject, while the second strategy will result in only one general pre-trained model. Under both strategies, we pre-train and validate our model on the Training and Validation Sets (i.e., repetitions (1, 3, 4, and 6) and repetitions (2 and 5) on Day 1), respectively. The calibration stage includes three strategies: ``0 Rep" calibration, ``1 Rep" calibration, and ``2 Reps" calibration. 0 Rep calibration directly evaluates a pre-trained model (either Pre-trained on Individuals or Pre-trained on All) on the Test Set (i.e., repetitions (2 and 5) on Day 2). To improve the performance of a pre-trained model on Day 2, we calibrate it on limited data from Day 2 by only retraining the Linear Projection layer (which accounts for 8.8\% of total trainable parameters) of the pre-trained model. We conduct 4-fold or 6-fold cross-validation for calibration by selecting one or two repetitions from the Calibration set (i.e., repetitions (1, 3, 4, and 6) on Day 2). The training/training-calibration strategy of model evaluation for intraday and interday performances can be visualized in Fig. \ref{fig:strategies}.

\subsection{Statistical Test}
\label{subsec:stats}
The distribution of the differences between two compared groups of results is not normal. Also, the compared groups of results are dependent. Thus, we use Wilcoxon signed-rank test with an alpha value of 0.05 in this study. Asterisks ($*$) are used to indicate statistical significance with respect to \textit{p} values. $ns$ or not significant denotes $p>0.05$; $*$ denotes $p \leq 0.05$; $**$ denotes $p \leq 0.01$; $***$ denotes $p \leq 0.001$; and $****$ denotes $p \leq 0.0001$.

\section{Results}
\label{sec:results}
\subsection{Model Configuration}
\label{subsec:model-config}
This paper particularly focuses on learning the relationships of signals from all the sensors at any timestamp of an HD-sEMG window to enhance the long-term, interday performance of HGR. Thus, the dimension of each patch ($patch\_size$) is the same as each electrode grid (i.e., $8 \times 8$). We consider $channels$ as the number of used electrode grids, which is four for the ``Hyser". As a result of hyper-parameter tuning, the dimension of the latent space $D$ is set to 128. The transformer encoder has eight layers ($L=8$), each having an MSA with four size 16 heads ($h=4$ and $d_k=16$). $mlp\_dim$, which is the dimension of the MLP block's hidden layer, is set to 32. The dropout rates after the position embedding and inside the transformer encoder are 0.1 and 0.5, respectively. The model configuration is summarized in Table \ref{tab:model-config}.

\begin{table}[ht]
    \caption{Model configuration (notation and value pairs).}
    \label{tab:model-config}
    \centering
    \setlength{\tabcolsep}{14pt}
    \begin{tabular}{|c|c||c|c|}
    \hline
        \textbf{Notation} & \textbf{Value} & \textbf{Notation} & \textbf{Value} \\
    \hline\hline
        $patch\_size$ & $8 \times 8$ & $channels$ & 4 \\
    \hline
        $D$ & 128 & $L$ & 8 \\
    \hline
        $h$ & 4 & $d_k$ & 16 \\
    \hline
        $mlp\_dim$ & 32 &  & \\
    \hline
    \end{tabular}
\end{table}

\subsection{Results of Evaluation Protocols}
\label{subsec:results-protocols}

\begin{figure}[tbh]
    \centering
    \includegraphics[width=\linewidth]{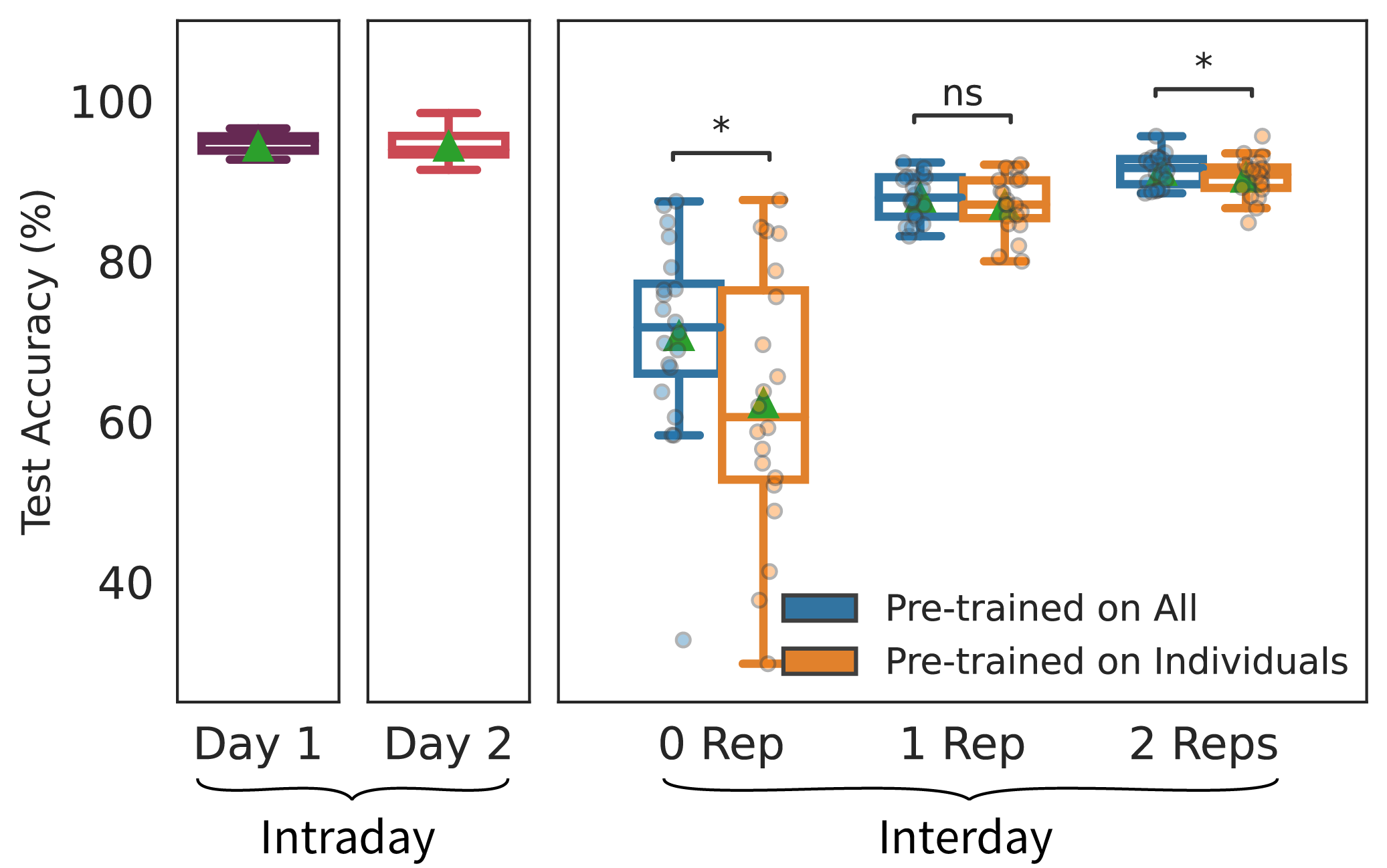}
    \caption{Box plots of intraday and interday performance given different pre-training strategies. The number of data points in each box plot is 20, equal to the total subjects. Green triangles denote the averages.}
    \label{fig:inter-2strategies}
\end{figure}

This paper aims to achieve high reliability and performance for day-to-day HGR without or with minimal calibration. We investigate the minimum amount of data needed for our proposed model pre-trained on Day 1 to maintain the performance on Day 2, lifting the data collection burden from users and enhancing the ease of use of such HGR systems. In this regard, we conduct 0 Rep calibration, 1 Rep calibration, and 2 Reps calibration strategies on sEMG signals collected on Day 2 using our pre-trained model (i.e., Pre-trained on Individuals or Pre-trained on All). The calibration data of the latter two strategies account for 17\% and 33\% of data collected on Day 2, respectively. As 1 Rep and 2 Reps calibrations are evaluated using 4-fold and 6-fold cross-validation, the average accuracies across folds are reported as the final results. Our proposed model's intraday and interday performances are shown in Table \ref{tab:inter-2strategies} and Fig. \ref{fig:inter-2strategies}.

\subsubsection*{Intraday performance} When training our proposed model from scratch on four repetitions (1, 3, 4, and 6), we can achieve consistently high accuracies of 94.32\%$\pm$2.66\% for Day 1 and 94.33\%$\pm$1.78\% for Day 2 averaged across 20 subjects. The distributions of sEMG signals collected on Day 1 and Day 2 are different due to sEMG variation over time, though the data collection followed the same protocol. Thus, the performance consistency shows that our proposed model performs stably in single-day HGR.

\begin{table}[htb]
    \caption{Average interday performance by pre-training strategy.}
    \label{tab:inter-2strategies}
    \centering
    \setlength{\tabcolsep}{10pt}
    \begin{tabular}{|c|c|c|c|}
        \hline
        \textbf{Pre-training Strategy} & \textbf{0 Rep} & \textbf{1 Rep} & \textbf{2 Reps} \\\hline\hline
        Pre-trained on All & \textbf{71.34\%} & \textbf{88.87\%} & \textbf{92.25\%} \\\hline
        Pre-trained on Individuals & 62.84\% & 87.92\% & 91.38\% \\\hline
    \end{tabular}
\end{table}

\subsubsection*{Interday performance} Pre-trained on Individuals models can achieve 62.84\%$\pm$16.67\%, 87.92\%$\pm$3.57\%, and 91.38\%$\pm$2.50\% averaged across all subjects in 0 Rep, 1 Rep, and 2 Reps calibrations. In comparison, these performances are 8.5\%$\pm$3.91\%, 0.95\%$\pm$0.76\%, and 0.87\%$\pm$0.5\% higher when using the Pre-trained on All model. It should be noted that the average standard deviations are lower, indicating that our model pre-trained on all subjects is more confident in predicting gestures based on sEMG signals collected on Day 2 than the Pre-trained on Individuals models. Thus, pre-training on all subjects can help our proposed model learn time-invariant HGR-related knowledge from sEMG signals collected on Day 1. This knowledge can be more generalized, emphasized, and strengthened when learned from all subjects than captured from individuals. Furthermore, the interday performance of our proposed model can almost match the intraday one when using the Pre-trained on All and 2 Reps calibration strategies only with a 2.08\% accuracy gap.

\begin{table}[htb]
    \caption{Average interday performance by window size under pre-trained on all strategy.}
    \label{tab:inter-win-size}
    \centering
    \setlength{\tabcolsep}{14pt}
    \begin{tabular}{|c|c|c|c|}
        \hline
        \textbf{Window Size} & \textbf{0 Rep} & \textbf{1 Rep} & \textbf{2 Reps} \\
        \hline\hline
        30 ms & 66.58\% & 86.19\% & 89.59\% \\\hline
        40 ms & 69.20\% & 86.81\% & 90.31\% \\\hline
        50 ms & 71.34\% & 88.87\% & 92.25\% \\\hline
        100 ms & 74.16\% & 89.65\% & 92.44\% \\\hline
        200 ms & 74.70\% & 91.92\% & 94.46\% \\\hline
    \end{tabular}
\end{table}

\begin{figure}[tbh]
    \centering
    \includegraphics[width=\linewidth]{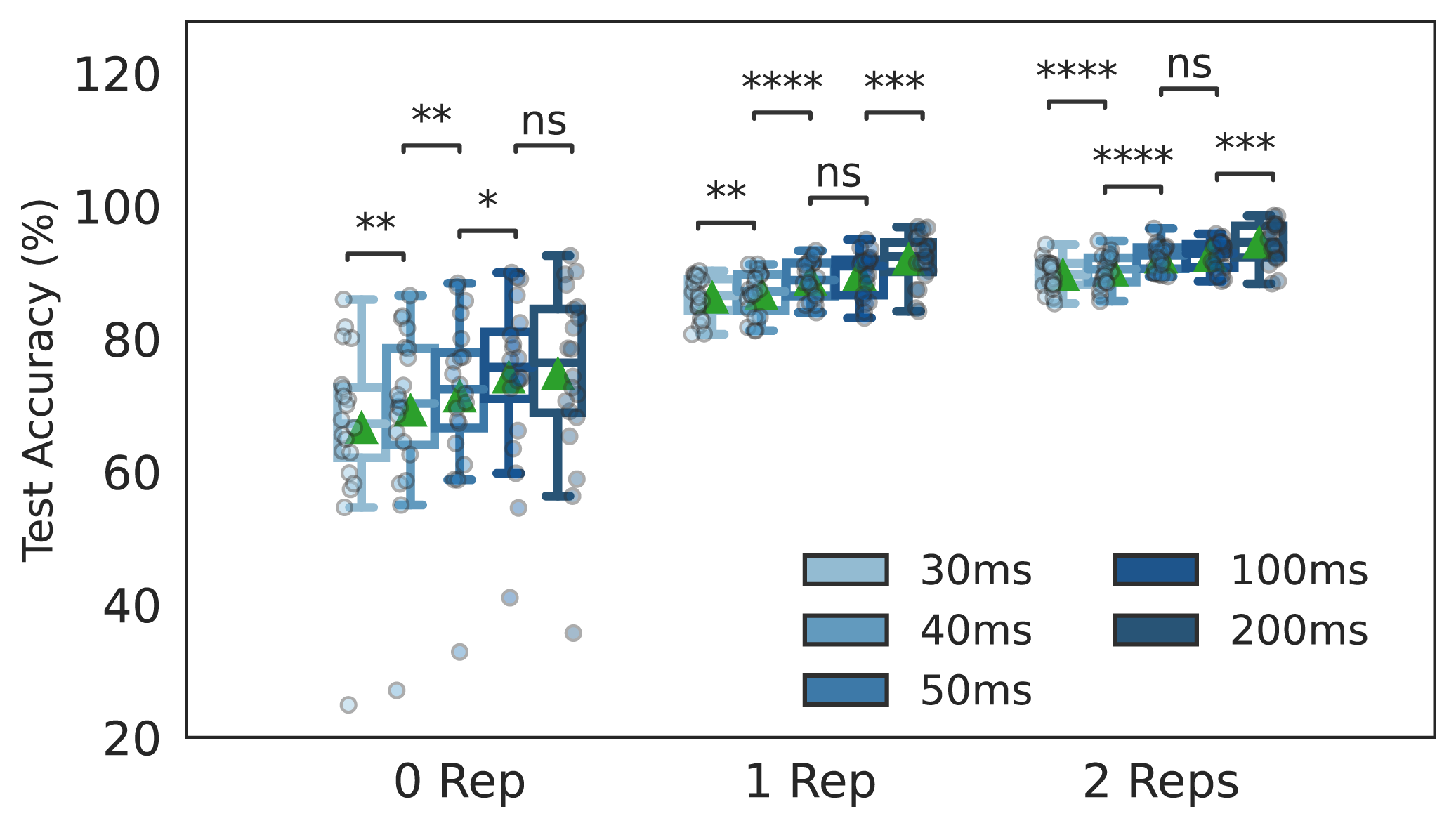}
    \caption{Box plots of interday performance by window size under Pre-trained on All strategy. The number of data points in each box plot is 20, equal to the total subjects. Green triangles denote the averages. The shade of color darkens as the window size increases.}
    \label{fig:inter-win}
\end{figure}

There is a trade-off between window size and the performance of an HGR system. Larger window sizes contain more information that can be important for the proposed model to learn to enhance HGR performance. However, larger window sizes can result in longer induced delays to myoelectric control and more complex model structures. This paper aims to find the shortest window size that can achieve the optimal interday HGR performance, preparing our proposed model for real-time cross-day use. Previous works \cite{Sun2022-ib,Montazerin2023-yr} achieved high intraday performance in classifying a large number of multi-degree-of-freedom gestures using short window sizes ranging from 30 ms to 250 ms. Hence, in this paper, we investigate the effect of window sizes on interday performance by pre-training (under the Pre-trained on All strategy) and calibrate our proposed model on five window sizes: 30 ms, 40 ms, 50 ms, 100 ms, and 200 ms. The results are shown in Table \ref{tab:inter-win-size} and Fig. \ref{fig:inter-win}. We statistically analyze and compare the interday performances on (30 ms, 40 ms), (40 ms, 50 ms), (50 ms, 100 ms), and (100 ms, 200 ms) pairs of window sizes on each calibration strategy. As shown in Fig. \ref{fig:inter-win}, the performance on 50 ms windows is significantly higher than on 30 ms and 40 ms windows but similar to the performance on 100 ms windows. As a result, our proposed model achieves the most optimistic interday performance on 50 ms windows.

\begin{figure}[tbh]
    \centering
    \includegraphics[width=\linewidth]{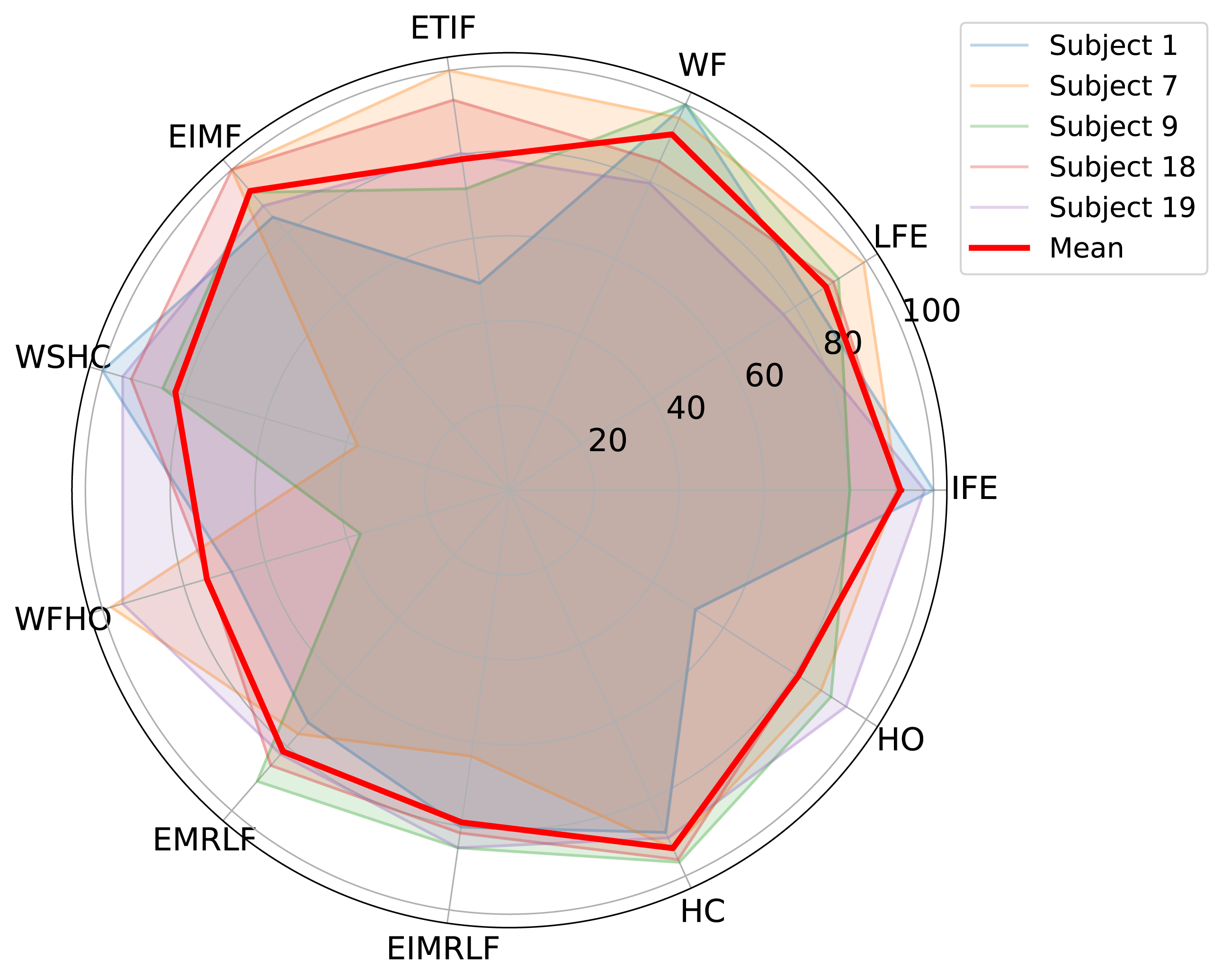}
    \caption{Gesture-wise average interday performance without calibration (0 Rep calibration) on the top-5-performing subjects. The bold red line denotes the interday performance averaged across the five subjects.}
    \label{fig:radar-top5}
\end{figure}

This paper also investigates gesture-wise reliability across days by analyzing the interday performance of our proposed model without calibration (0 Rep calibration) for each gesture performed by the top-5-performing subjects (subjects 1, 7, 9, 18, and 19). As a result, Fig. \ref{fig:radar-top5} shows that our proposed model can robustly and reliably predict almost all gestures on two different days. Our model achieves more than 90\% accuracy in predicting five gestures (IFE, WF, EIMF, HO, and EIMRLF). Only one gesture (WSCHC) has an accuracy of less than 75\%.

\section{Comparative Study}
\label{sec:comparative}
 In this study, we present a novel strategy to generalize HGR across days while requiring minimal calibration. In order to evaluate the performance of our model compared to the state of the art, we comprehensively conduct three sets of comparative experiments. In one experiment, we compare our model with one of the most common approaches toward generalizable gesture detection using sEMG signals: feature engineering combined with ALDA. We show that our model significantly outperforms ALDA under similar training and calibration configurations. To further highlight the superiority of our proposed model over other state-of-the-art models, in another experiment, we compare the performance of our ViT-MDHGR with three well-known RNN models which have proven to be effective when addressing time-series data \cite{Quivira2018-ls}. Finally, as three-dimensional CNNs (3D CNNs) \cite{Sun2022cnn} become increasingly popular in HGR for extracting temporal and spatial information from HD-sEMG using 3D kernels, we compare our proposed model with a 3D CNN model.
 These experiments follow the same pre-training (i.e., Pre-trained on All) and calibration strategies as our ViT-MDHGR. The experimental details are described as follows.

 \subsubsection*{Comparison to ALDA} ALDA based on feature engineering has been subject to extensive studies in the gesture recognition literature \cite{Jiang2018-rl,Botros2022-stability,Hohne2015-covariate}. In order to demonstrate the superiority of our ViT-MDHGR over ALDA, we adopt a feature engineering + ALDA model to the same dataset (``Hyser") following the general methodology of \cite{Jiang2018-rl}. More specifically, we create a pipeline that extracts a number of features from each 50 ms signal window and classifies the feature vectors into gestures using an LDA classifier. The features we considered are Mean Absolute Value, Zero Crossing, Slope Sign Changes, and Waveform Length similar to \cite{Jiang2018-rl}. The training and calibration stages of ALDA are described as follows.
 \begin{itemize}
     \item \textbf{Training:} Features are extracted from the Training Set of all 20 subjects. Next, an LDA with the following discriminant function is fit to the extracted features following \cite{Jiang2018-rl,Hohne2015-covariate}:
     \begin{equation}
         g_c(x) = x^\intercal \Sigma^{-1} \mu_c - \frac{1}{2}\mu_c^\intercal \Sigma^{-1}\mu_c + \log \pi_c
     \end{equation}
     where $\mu_c$ is the mean value for class $c$, $\Sigma$ is the covariance matrix of classes (assumed to be equal for all classes), and $\pi_c$ is the prior probability for class $c$.
     \item \textbf{Calibration:} Given new samples on Day 2 (Any one or two repetitions selected from the Calibration Set) for each subject, we adapt the LDA by readjusting the mean vectors and covariance matrix similar to \cite{Hohne2015-covariate} as follows:

    \begin{subequations}
        \begin{align}
        \tilde{\mu}_c &= (1-\lambda) \mu_{tr_c}+\lambda \mu_{cal_c}\\
        \tilde{\Sigma}_c &= (1-\lambda) \Sigma_{tr_c}+\lambda \Sigma_{cal_c}
        \end{align}
    \end{subequations}
     
     where $(\mu_{tr_c},\Sigma_{tr_c})$ refer to training (mean, covariance) and $(\mu_{cal_c},\Sigma_{cal_c})$ refer to calibration data (mean, covariance). $\lambda=0.5$ is a hyperparameter.\\
 \end{itemize}
  
\subsubsection*{Comparison to RNNs}
The compared models in this experiment set are Gated Recurrent Units (GRUs)\cite{Chung2014-gr}, Long Short Term Memories (LSTMs) \cite{Hochreiter1997-ls}, and Bidirectional LSTMs (BILSTMs). Each model has three RNN layers followed by FC layer(s) with hyperbolic tangent activation and Dropout (rate=0.2) layers. The last layer is an FC layer with softmax activation that outputs a predicted gesture. In the calibration stage, only the first RNN layer is trainable. Table \ref{tab:rnn-config} shows the configurations and hyperparameters of RNN models.

\begin{table}[ht]
    \caption{RNN Model configurations}
    \label{tab:rnn-config}
    \centering
    \setlength{\tabcolsep}{12pt}
    \begin{tabular}{|c|c|c|c|}
    \hline
        \textbf{Hyperparameter} & \textbf{GRU} & \textbf{LSTM} & \textbf{BILSTM}\\
    \hline\hline
        layers & $3$ &  $3$  & $3$ \\
    \hline
        hidden units & $134$ & $115$ & $75$\\
    \hline
        dilation order & $3$ & $3$ & $3$ \\
    \hline
        FC layers before classifier & 2 & 1 & 1 \\
    \hline 
        Dropout & None & 1 & 1 \\
    \hline 
    \end{tabular}
\end{table}

\subsubsection*{Comparison to 3D CNN} Similar to the previous experiments, we compare the ViT-MDHGR model with a 3D CNN model structured as shown in Table \ref{tab:3dcnn}. For calibration, we freeze everything except the second 3D convolution and batch normalization layers.

\begin{table}[ht]
    \caption{Model structure for 3D CNN. c: channels/units, k: kernel size, d: dilation rate, a: activation.}
    \label{tab:3dcnn}
    \centering
    \setlength{\tabcolsep}{12pt}
    \begin{tabular}{|c|c|}
    \hline
        \textbf{Layer} & \textbf{Configuration} \\
    \hline\hline
    3D convolution & c = 16, k = (8, 2, 2), d=(2, 1, 1), a='relu' \\
    \hline
    Batch normalization & N/A \\
    \hline
    3D convolution & c=32, k=(8, 2, 2), d=(2, 1, 1), a='relu' \\
    \hline
    Batch normalization & N/A \\
    \hline
    3D Max pooling & N/A \\
    \hline
    3D convolution & c=64, k=(8, 2, 2), d=(4, 1, 1), a='relu' \\
    \hline
    Batch normalization & N/A \\
    \hline
    3D convolution & c=64, k=(8, 2, 2), d=N/A,  a='relu' \\
    Batch normalization & N/A \\
    \hline
    flatten & N/A \\
    \hline
    FC & c=32, a='tanh'\\
    \hline
    Dropout & rate=0.2 \\
    \hline
    FC (classifier) & c=11, a='softmax' \\
    \hline
    \end{tabular}
\end{table}

\begin{table}[ht]
    \caption{Trainable parameters in pre-training and calibration stages.}
    \label{tab:params}
    \centering\setlength{\tabcolsep}{10pt}
    \begin{tabular}{|c|c|c|c|}
    \hline
        \textbf{Model} & \textbf{Pre-training} & \textbf{Calibration} &\textbf{Percentage} \\
    \hline \hline
        ViT-MDHGR & 382,475 & 33,664 & 8.8\%\\
    \hline
        GRU &  385,747 & 157,584 & 40.8\% \\
    \hline
        LSTM & 387,715 & 171,120 & 44.1\% \\
    \hline
        BILSTM & 385,595 & 199,200 & 51.6\% \\
    \hline
        3D CNN & 218,011 & 16,480 & 7.5\% \\
    \hline
    \end{tabular}
\end{table}

\begin{table}[ht]
    \caption{Performance evaluation of different models. Mean values are reported for calibration results.}
    \label{tab:compare-acc}
    \centering
    \setlength{\tabcolsep}{12pt}
    \begin{tabular}{|c|c|c|c|}
    \hline
        \textbf{Model} & \textbf{0 Rep} & \textbf{1 Rep} & \textbf{2 Reps} \\
    \hline\hline
        ViT-MDHGR & \textbf{71.34\%} & \textbf{88.87\%} & \textbf{92.25\%} \\
    \hline
        GRU &  69.02\% & 78.07\% & 81.20\% \\
    \hline
        LSTM &  70.26\% & 81.41\% & 84.59\% \\
    \hline
        BILSTM &  70.61\% & 82.11\% & 84.80\% \\
    \hline
        ALDA &  50.84\% & 74.76\% &  81.41\%\\
    \hline
        3D CNN &  67.49\% & 79.47\% & 83.05\%\\
    \hline
    \end{tabular}
\end{table}

\begin{figure}
    \centering
    \includegraphics[width=\linewidth]{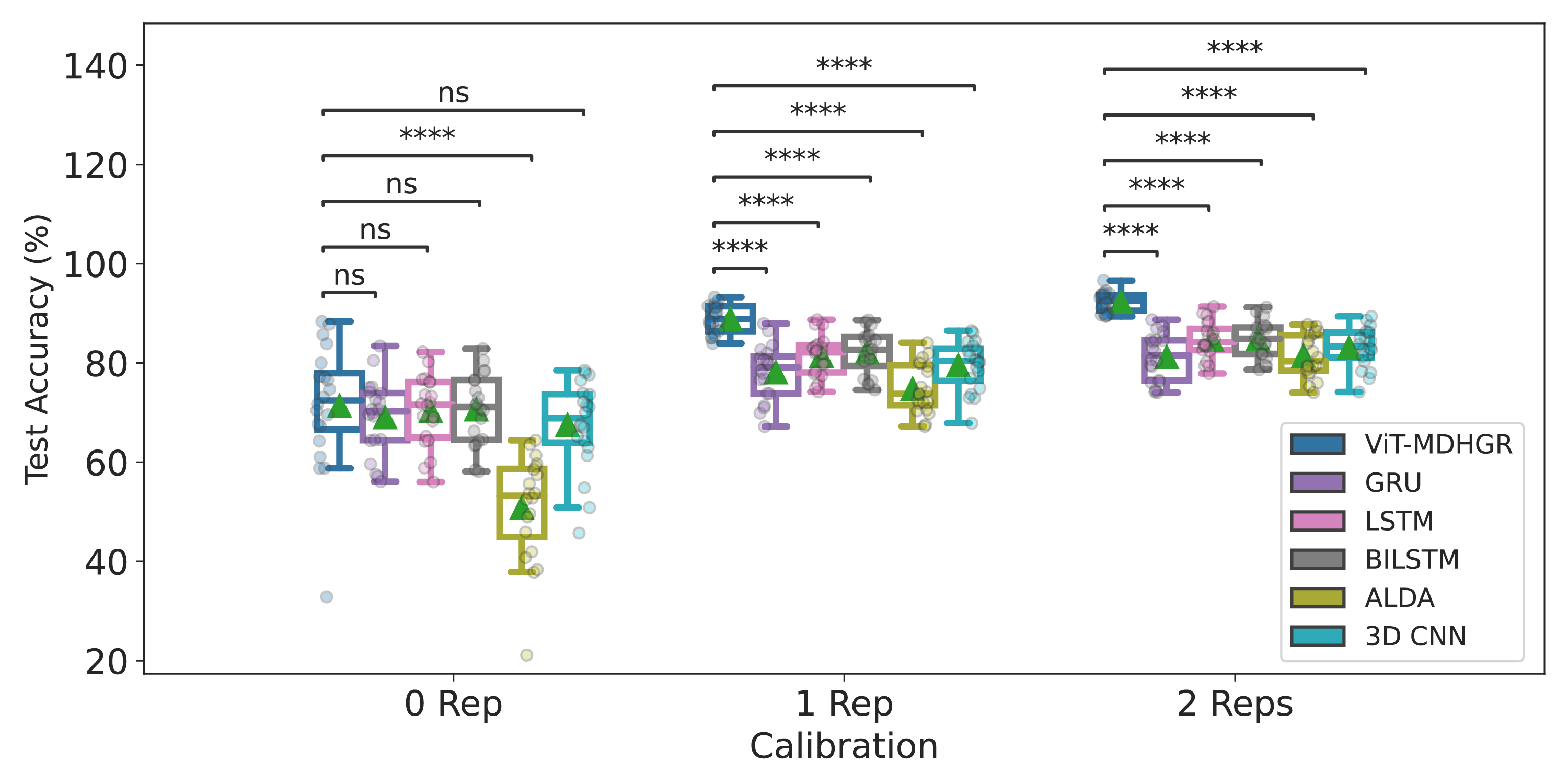}
    \caption{Day 2 Accuracy comparison between ViT-MDHGR and various state of the art models. The number of data points in each box plot is 20, equal to the total subjects. Green triangles show mean values across 20 subjects.}
    \label{fig:comparison}
\end{figure}

Table \ref{tab:params} shows the total number of trainable parameters in the pre-training and calibration stages for each of the models. The ratio of the trainable parameters of the calibration stage to the pre-training stage is also shown. We note that ViT-MDHGR and 3D CNN require significantly lower ratios of trainable parameters during calibration. Table \ref{tab:compare-acc} shows how these models compare to each other in various interday calibration scenarios. Mean values are shown for various Calibration percentages and the models are calibrated on individual subjects separately. Fig. \ref{fig:comparison} gives a more detailed view of Calibration results for the two models. We make the following observations:\\

\textit{Observation 1:} ViT-MDHGR slightly outperforms the RNN models when evaluated on Day 2 data with 0 Rep calibration. However, by calibrating over one repetition, the performance gap increases to more than $6\%$. Calibrating on two repetitions increases this gap to around $8\%$ compared to the best RNN model. We emphasize that in all scenarios, our proposed model outperforms RNN models while requiring calibration over significantly fewer parameters.\\

\textit{Observation 2:} ViT-MDHGR significantly outperforms ALDA in all scenarios. When calibrating on two repetitions, this gap is around $10.84\%$. In comparison to ALDA, We emphasize that our proposed model does not require any additional pre-processing or feature extraction. ViT-MDHGR also outperforms 3D CNN, more significantly when two repetitions for calibration are available.\\

\textit{Observation 3:} When calibrated on one or two repetitions, ViT-MDHGR is more reliable in the sense that it has a lower variance among 20 subjects. We also note that when calibrated on two repetitions, the worst ViT-MDHGR accuracy is still higher than the third quartile values of all the other models.\\

In general, these results confirm the potential and reliability of our proposed model for accurate and agile gesture recognition across days.

\section{Conclusion}
\label{sec:conclusion}
In this study, we propose a ViT-based network (ViT-MDHGR) that can be deployed for real-time HGR from HD-sEMG signals. Our proposed model addresses the challenge of inter-day gesture recognition relying on 50 ms HD-sEMG signal windows. This innovation enhances the agility and responsiveness of the system, making it suitable for practical applications. Retraining only $8.8\%$ of the model parameters on a different day, we show that this model can achieve an inter-day accuracy of $92.25\%$ for detecting 11 gestures by calibration on only two repetitions of each gesture. We demonstrate that our proposed model significantly outperforms various sequential, CNN, and LDA based state of the art networks.
This research highlights a significant step towards making multi-day hand gesture recognition a reality, with potential applications in myoelectric control of prosthetic devices, neurorobotics, and human-computer interfaces. The ViT-MDHGR not only improves the generalizability of hand gesture recognition but also offers a promising avenue for enhancing the usability and practicality of such systems in real-world contexts. 
Possible directions of future work include extending the number of gestures, incorporating subject-generalization, and increasing the length of inter-day intervals. Focusing on refining and expanding the capabilities of this model will bring us closer to seamless and efficient myoelectric control solutions for a wide range of applications.

\section*{Acknowledgment}
We would like to acknowledge Mohammedali Roowala's contributions to this paper. Roowala is with the Department of Mechanical and Aerospace Engineering, New York University (NYU), New York, NY, 11201 USA.

\section*{References}
\vspace{-2em}
\bibliographystyle{IEEEtran}
\bibliography{bibliography}

\begin{thebibliography}{10}
\providecommand{\url}[1]{#1}
\csname url@samestyle\endcsname
\providecommand{\newblock}{\relax}
\providecommand{\bibinfo}[2]{#2}
\providecommand{\BIBentrySTDinterwordspacing}{\spaceskip=0pt\relax}
\providecommand{\BIBentryALTinterwordstretchfactor}{4}
\providecommand{\BIBentryALTinterwordspacing}{\spaceskip=\fontdimen2\font plus
\BIBentryALTinterwordstretchfactor\fontdimen3\font minus \fontdimen4\font\relax}
\providecommand{\BIBforeignlanguage}[2]{{%
\expandafter\ifx\csname l@#1\endcsname\relax
\typeout{** WARNING: IEEEtran.bst: No hyphenation pattern has been}%
\typeout{** loaded for the language `#1'. Using the pattern for}%
\typeout{** the default language instead.}%
\else
\language=\csname l@#1\endcsname
\fi
#2}}
\providecommand{\BIBdecl}{\relax}
\BIBdecl

\bibitem{Feher2012-ia}
J.~Feher, ``3.6 - the neuromuscular junction and {Excitation-Contraction} coupling,'' in \emph{Quantitative Human Physiology}, J.~Feher, Ed.\hskip 1em plus 0.5em minus 0.4em\relax Boston: Academic Press, Jan. 2012, pp. 259--269.

\bibitem{Del_Vecchio2020-if}
A.~Del~Vecchio, A.~Holobar, D.~Falla, F.~Felici, R.~M. Enoka, and D.~Farina, ``\BIBforeignlanguage{en}{Tutorial: Analysis of motor unit discharge characteristics from high-density surface {EMG} signals},'' \emph{\BIBforeignlanguage{en}{J. Electromyogr. Kinesiol.}}, vol.~53, p. 102426, Aug. 2020.

\bibitem{Raez2006-ro}
M.~B.~I. Raez, M.~S. Hussain, and F.~Mohd-Yasin, ``\BIBforeignlanguage{en}{Techniques of {EMG} signal analysis: detection, processing, classification and applications},'' \emph{\BIBforeignlanguage{en}{Biol. Proced. Online}}, vol.~8, pp. 11--35, Mar. 2006.

\bibitem{Bi2019-up}
L.~Bi, A.-g. Feleke, and C.~Guan, ``A review on {EMG-based} motor intention prediction of continuous human upper limb motion for human-robot collaboration,'' \emph{Biomed. Signal Process. Control}, vol.~51, pp. 113--127, May 2019.

\bibitem{Zhang2022-ch}
Y.~Zhang, B.~Liang, B.~Chen, P.~M. Torrens, S.~F. Atashzar, D.~Lin, and Q.~Sun, ``{Force-Aware} interface via electromyography for natural {VR/AR} interaction,'' \emph{ACM Trans. Graph.}, vol.~41, no.~6, pp. 1--18, Nov. 2022.

\bibitem{Ergeneci2022-eq}
M.~Ergeneci, D.~Carter, and P.~Kosmas, ``{sEMG} onset detection via bidirectional recurrent neural networks with applications to sports science,'' \emph{IEEE Sens. J.}, vol.~22, no.~19, pp. 18\,751--18\,761, Oct. 2022.

\bibitem{Kumar2022-ux}
D.~Kumar and A.~Ganesh, ``\BIBforeignlanguage{en}{A critical review on hand gesture recognition using {sEMG}: Challenges, application, process and techniques},'' \emph{\BIBforeignlanguage{en}{J. Phys. Conf. Ser.}}, vol. 2327, no.~1, p. 012075, Aug. 2022.

\bibitem{Shehata2021-ns}
A.~W. Shehata, H.~E. Williams, J.~S. Hebert, and P.~M. Pilarski, ``Machine learning for the control of prosthetic arms: Using electromyographic signals for improved performance,'' \emph{IEEE Signal Process. Mag.}, vol.~38, no.~4, pp. 46--53, Jul. 2021.

\bibitem{Holobar2021-hb}
A.~Holobar and D.~Farina, ``Noninvasive neural interfacing with wearable muscle sensors: Combining convolutive blind source separation methods and deep learning techniques for neural decoding,'' \emph{IEEE Signal Process. Mag.}, vol.~38, no.~4, pp. 103--118, Jul. 2021.

\bibitem{Chowdhury2013-wf}
R.~H. Chowdhury, M.~B.~I. Reaz, M.~A. B.~M. Ali, A.~A.~A. Bakar, K.~Chellappan, and T.~G. Chang, ``\BIBforeignlanguage{en}{Surface electromyography signal processing and classification techniques},'' \emph{\BIBforeignlanguage{en}{Sensors}}, vol.~13, no.~9, pp. 12\,431--12\,466, Sep. 2013.

\bibitem{Botros2022-rd}
F.~S. Botros, A.~Phinyomark, and E.~J. Scheme, ``{Day-to-Day} stability of wrist {EMG} for {Wearable-Based} hand gesture recognition,'' \emph{IEEE Access}, vol.~10, pp. 125\,942--125\,954, 2022.

\bibitem{Englehart2003-iq}
K.~Englehart and B.~Hudgins, ``\BIBforeignlanguage{en}{A robust, real-time control scheme for multifunction myoelectric control},'' \emph{\BIBforeignlanguage{en}{IEEE Trans. Biomed. Eng.}}, vol.~50, no.~7, pp. 848--854, Jul. 2003.

\bibitem{Zia_ur_Rehman2018-mk}
M.~Zia~ur Rehman, S.~O. Gilani, A.~Waris, I.~K. Niazi, G.~Slabaugh, D.~Farina, and E.~N. Kamavuako, ``\BIBforeignlanguage{en}{Stacked sparse autoencoders for {EMG-Based} classification of hand motions: A comparative multi day analyses between surface and intramuscular {EMG}},'' \emph{\BIBforeignlanguage{en}{NATO Adv. Sci. Inst. Ser. E Appl. Sci.}}, vol.~8, no.~7, p. 1126, Jul. 2018.

\bibitem{Wang2019-qd}
Z.~Wang, Y.~Fang, G.~Li, and H.~Liu, ``Facilitate {sEMG-Based} {Human--Machine} interaction through channel optimization,'' \emph{Int. J. Humanoid Rob.}, vol.~16, no.~04, p. 1941001, Aug. 2019.

\bibitem{Meng2022-hn}
L.~Meng, X.~Jiang, X.~Liu, J.~Fan, H.~Ren, Y.~Guo, H.~Diao, Z.~Wang, C.~Chen, C.~Dai, and W.~Chen, ``{User-Tailored} hand gesture recognition system for wearable prosthesis and armband based on surface electromyogram,'' \emph{IEEE Trans. Instrum. Meas.}, vol.~71, pp. 1--16, 2022.

\bibitem{Jiang2022-di}
X.~Jiang, X.~Liu, J.~Fan, X.~Ye, C.~Dai, E.~A. Clancy, D.~Farina, and W.~Chen, ``Optimization of {HD-sEMG-Based} {Cross-Day} hand gesture classification by optimal feature extraction and data augmentation,'' \emph{IEEE Transactions on Human-Machine Systems}, vol.~52, no.~6, pp. 1281--1291, Dec. 2022.

\bibitem{Waris2019-jx}
A.~Waris, I.~K. Niazi, M.~Jamil, K.~Englehart, W.~Jensen, and E.~N. Kamavuako, ``\BIBforeignlanguage{en}{Multiday evaluation of techniques for {EMG-Based} classification of hand motions},'' \emph{\BIBforeignlanguage{en}{IEEE J Biomed Health Inform}}, vol.~23, no.~4, pp. 1526--1534, Jul. 2019.

\bibitem{Fang2021-ro}
Y.~Fang, X.~Zhang, D.~Zhou, and H.~Liu, ``Improve inter-day hand gesture recognition via convolutional neural network-based feature fusion,'' \emph{Int. J. Humanoid Rob.}, vol.~18, no.~02, p. 2050025, Apr. 2021.

\bibitem{Zanghieri2020-tf}
M.~Zanghieri, S.~Benatti, F.~Conti, A.~Burrello, and L.~Benini, ``Temporal variability analysis in {sEMG} hand grasp recognition using temporal convolutional networks,'' in \emph{2020 2nd {IEEE} International Conference on Artificial Intelligence Circuits and Systems ({AICAS})}, Aug. 2020, pp. 228--232.

\bibitem{Qureshi2022-wl}
M.~F. Qureshi, Z.~Mushtaq, M.~Z.~u. Rehman, and E.~N. Kamavuako, ``Spectral {Image-Based} multiday surface electromyography classification of hand motions using {CNN} for {Human--Computer} interaction,'' \emph{IEEE Sens. J.}, vol.~22, no.~21, pp. 20\,676--20\,683, Nov. 2022.

\bibitem{Ketyko2019-rm}
\emph{Domain Adaptation for {sEMG-based} Gesture Recognition with Recurrent Neural Networks}, Jan. 2019.

\bibitem{Du2017-il}
Y.~Du, W.~Jin, W.~Wei, Y.~Hu, and W.~Geng, ``\BIBforeignlanguage{en}{Surface {EMG-Based} {Inter-Session} gesture recognition enhanced by deep domain adaptation},'' \emph{\BIBforeignlanguage{en}{Sensors}}, vol.~17, no.~3, Feb. 2017.

\bibitem{Cote-Allard2020-nr}
U.~C{\^o}t{\'e}-Allard, G.~Gagnon-Turcotte, A.~Phinyomark, K.~Glette, E.~J. Scheme, F.~Laviolette, and B.~Gosselin, ``Unsupervised domain adversarial {Self-Calibration} for {Electromyography-Based} gesture recognition,'' \emph{IEEE Access}, vol.~8, pp. 177\,941--177\,955, 2020.

\bibitem{Fang2017-fd}
Y.~Fang, D.~Zhou, K.~Li, and H.~Liu, ``\BIBforeignlanguage{en}{Interface prostheses with {Classifier-Feedback-Based} user training},'' \emph{\BIBforeignlanguage{en}{IEEE Trans. Biomed. Eng.}}, vol.~64, no.~11, pp. 2575--2583, Nov. 2017.

\bibitem{Raurale2020-ma}
S.~A. Raurale, J.~McAllister, and J.~M. del Rincon, ``{Real-Time} embedded {EMG} signal analysis for {Wrist-Hand} pose identification,'' \emph{IEEE Trans. Signal Process.}, vol.~68, pp. 2713--2723, 2020.

\bibitem{Tavakoli2018-hb}
M.~Tavakoli, C.~Benussi, P.~Alhais~Lopes, L.~B. Osorio, and A.~T. de~Almeida, ``Robust hand gesture recognition with a double channel surface {EMG} wearable armband and {SVM} classifier,'' \emph{Biomed. Signal Process. Control}, vol.~46, pp. 121--130, Sep. 2018.

\bibitem{Amsuess2016-hp}
S.~Amsuess, I.~Vujaklija, P.~Goebel, A.~D. Roche, B.~Graimann, O.~C. Aszmann, and D.~Farina, ``\BIBforeignlanguage{en}{{Context-Dependent} upper limb prosthesis control for natural and robust use},'' \emph{\BIBforeignlanguage{en}{IEEE Trans. Neural Syst. Rehabil. Eng.}}, vol.~24, no.~7, pp. 744--753, Jul. 2016.

\bibitem{Lee2020-tb}
S.~Lee, B.~Jamil, S.~Kim, and Y.~Choi, ``\BIBforeignlanguage{en}{Fabric vest socket with embroidered electrodes for control of myoelectric prosthesis},'' \emph{\BIBforeignlanguage{en}{Sensors}}, vol.~20, no.~4, Feb. 2020.

\bibitem{Sun2022-ib}
T.~Sun, Q.~Hu, J.~Libby, and S.~F. Atashzar, ``Deep heterogeneous dilation of {LSTM} for {Transient-Phase} gesture prediction through {High-Density} electromyography: Towards application in neurorobotics,'' \emph{IEEE Robotics and Automation Letters}, vol.~7, no.~2, pp. 2851--2858, Apr. 2022.

\bibitem{Liu2023-mc}
Y.~Liu, X.~Li, L.~Yang, G.~Bian, and H.~Yu, ``A {CNN-Transformer} hybrid recognition approach for {sEMG-Based} dynamic gesture prediction,'' \emph{IEEE Trans. Instrum. Meas.}, vol.~72, pp. 1--16, 2023.

\bibitem{Stachaczyk2020-je}
M.~Stachaczyk, S.~F. Atashzar, and D.~Farina, ``\BIBforeignlanguage{en}{Adaptive spatial filtering of {High-Density} {EMG} for reducing the influence of noise and artefacts in myoelectric control},'' \emph{\BIBforeignlanguage{en}{IEEE Trans. Neural Syst. Rehabil. Eng.}}, vol.~28, no.~7, pp. 1511--1517, Jul. 2020.

\bibitem{Boschmann2014-nc}
A.~Boschmann and M.~Platzner, ``Towards robust {HD} {EMG} pattern recognition: Reducing electrode displacement effect using structural similarity,'' in \emph{2014 36th Annual International Conference of the {IEEE} Engineering in Medicine and Biology Society}, Aug. 2014, pp. 4547--4550.

\bibitem{Moin2018-xb}
A.~Moin, A.~Zhou, A.~Rahimi, S.~Benatti, A.~Menon, S.~Tamakloe, J.~Ting, N.~Yamamoto, Y.~Khan, F.~Burghardt, L.~Benini, A.~C. Arias, and J.~M. Rabaey, ``An {EMG} gesture recognition system with flexible {High-Density} sensors and {Brain-Inspired} {High-Dimensional} classifier,'' in \emph{2018 {IEEE} International Symposium on Circuits and Systems ({ISCAS})}, May 2018, pp. 1--5.

\bibitem{Phinyomark2020-jb}
A.~Phinyomark, E.~Campbell, and E.~Scheme, ``Surface electromyography ({EMG}) signal processing, classification, and practical considerations,'' in \emph{Biomedical Signal Processing: Advances in Theory, Algorithms and Applications}, G.~Naik, Ed.\hskip 1em plus 0.5em minus 0.4em\relax Singapore: Springer Singapore, 2020, pp. 3--29.

\bibitem{Jiang2023-sd}
N.~Jiang, C.~Chen, J.~He, J.~Meng, L.~Pan, S.~Su, and X.~Zhu, ``\BIBforeignlanguage{en}{Bio-robotics research for non-invasive myoelectric neural interfaces for upper-limb prosthetic control: a 10-year perspective review},'' \emph{\BIBforeignlanguage{en}{Natl Sci Rev}}, vol.~10, no.~5, p. nwad048, May 2023.

\bibitem{Prakash2021-gq}
A.~Prakash and S.~Sharma, ``\BIBforeignlanguage{en}{A low-cost transradial prosthesis controlled by the intention of muscular contraction},'' \emph{\BIBforeignlanguage{en}{Australas. Phys. Eng. Sci. Med.}}, vol.~44, no.~1, pp. 229--241, Mar. 2021.

\bibitem{Hudgins1993-rq}
B.~Hudgins, P.~Parker, and R.~N. Scott, ``\BIBforeignlanguage{en}{A new strategy for multifunction myoelectric control},'' \emph{\BIBforeignlanguage{en}{IEEE Trans. Biomed. Eng.}}, vol.~40, no.~1, pp. 82--94, Jan. 1993.

\bibitem{Gu2018-oo}
Y.~Gu, D.~Yang, Q.~Huang, W.~Yang, and H.~Liu, ``Robust {EMG} pattern recognition in the presence of confounding factors,'' \emph{Expert Syst. Appl.}, vol.~96, no.~C, pp. 208--217, Apr. 2018.

\bibitem{Jiang2018-rl}
S.~Jiang, B.~Lv, W.~Guo, C.~Zhang, H.~Wang, X.~Sheng, and P.~B. Shull, ``Feasibility of wrist-worn, real-time hand, and surface gesture recognition via semg and imu sensing,'' \emph{IEEE Transactions on Industrial Informatics}, vol.~14, no.~8, pp. 3376--3385, 2018.

\bibitem{Botros2022-stability}
F.~S. Botros, A.~Phinyomark, and E.~J. Scheme, ``Day-to-day stability of wrist emg for wearable-based hand gesture recognition,'' \emph{IEEE Access}, vol.~10, pp. 125\,942--125\,954, 2022.

\bibitem{Hohne2015-covariate}
M.~Höhne, h.-j. Hwang, S.~Amsuss, J.~Hahne, D.~Farina, and K.-R. Müller, ``Improving the robustness of myoelectric pattern recognition for upper limb prostheses by covariate shift adaptation,'' \emph{IEEE Transactions on Neural Systems and Rehabilitation Engineering}, vol.~24, pp. 1--1, 01 2015.

\bibitem{Zia_Ur_Rehman2018-bp}
M.~Zia Ur~Rehman, A.~Waris, S.~O. Gilani, M.~Jochumsen, I.~K. Niazi, M.~Jamil, D.~Farina, and E.~N. Kamavuako, ``\BIBforeignlanguage{en}{Multiday {EMG-Based} classification of hand motions with deep learning techniques},'' \emph{\BIBforeignlanguage{en}{Sensors}}, vol.~18, no.~8, Aug. 2018.

\bibitem{Wu2023-cs}
D.~Wu, J.~Yang, and M.~Sawan, ``\BIBforeignlanguage{en}{Transfer learning on electromyography ({EMG}) tasks: Approaches and beyond},'' \emph{\BIBforeignlanguage{en}{IEEE Trans. Neural Syst. Rehabil. Eng.}}, vol.~31, pp. 3015--3034, Jul. 2023.

\bibitem{Jiang2021-hy}
X.~Jiang, X.~Liu, J.~Fan, X.~Ye, C.~Dai, E.~A. Clancy, M.~Akay, and W.~Chen, ``Open access dataset, toolbox and benchmark processing results of high-density surface electromyogram recordings,'' \emph{IEEE Transactions on Neural Systems and Rehabilitation Engineering}, vol.~29, pp. 1035--1046, 2021.

\bibitem{dosovitskiy20}
A.~Dosovitskiy, L.~Beyer, A.~Kolesnikov, D.~Weissenborn, X.~Zhai, T.~Unterthiner, M.~Dehghani, M.~Minderer, G.~Heigold, S.~Gelly, J.~Uszkoreit, and N.~Houlsby, ``An image is worth 16x16 words: Transformers for image recognition at scale,'' \emph{arXiv [cs.CV]}, Oct. 2020.

\bibitem{Arnab2021-et}
\emph{{ViViT}: A Video Vision Transformer}, Mar. 2021.

\bibitem{Montazerin2023-yr}
M.~Montazerin, E.~Rahimian, F.~Naderkhani, S.~F. Atashzar, S.~Yanushkevich, and A.~Mohammadi, ``\BIBforeignlanguage{en}{Transformer-based hand gesture recognition from instantaneous to fused neural decomposition of high-density {EMG} signals},'' \emph{\BIBforeignlanguage{en}{Sci. Rep.}}, vol.~13, no.~1, p. 11000, Jul. 2023.

\bibitem{Vaswani2017-bv}
\emph{Attention Is All You Need}, Jun. 2017.

\bibitem{Quivira2018-ls}
F.~Quivira, T.~Koike-Akino, Y.~Wang, and D.~Erdogmus, ``Translating semg signals to continuous hand poses using recurrent neural networks,'' in \emph{2018 IEEE EMBS International Conference on Biomedical \& Health Informatics (BHI)}, 2018, pp. 166--169.

\bibitem{Sun2022cnn}
T.~Sun, J.~Libby, J.~Rizzo, and S.~F. Atashzar, ``Deep augmentation for electrode shift compensation in transient high-density semg: Towards application in neurorobotics,'' in \emph{2022 IEEE/RSJ International Conference on Intelligent Robots and Systems (IROS)}, 2022, pp. 6148--6153.

\bibitem{Chung2014-gr}
\BIBentryALTinterwordspacing
J.~Chung, {\c{C}}.~G{\"{u}}l{\c{c}}ehre, K.~Cho, and Y.~Bengio, ``Empirical evaluation of gated recurrent neural networks on sequence modeling,'' \emph{CoRR}, vol. abs/1412.3555, 2014. [Online]. Available: \url{http://arxiv.org/abs/1412.3555}
\BIBentrySTDinterwordspacing

\bibitem{Hochreiter1997-ls}
\BIBentryALTinterwordspacing
S.~Hochreiter and J.~Schmidhuber, ``{Long Short-Term Memory},'' \emph{Neural Computation}, vol.~9, no.~8, pp. 1735--1780, 11 1997. [Online]. Available: \url{https://doi.org/10.1162/neco.1997.9.8.1735}
\BIBentrySTDinterwordspacing

\end{thebibliography}

\end{document}